\documentclass[journal]{IEEEtran}

\usepackage{subfigure}
\usepackage{setspace}
\usepackage{multirow} 
\usepackage{algorithm}
\usepackage{algorithmic}
\usepackage{amsmath}
\usepackage{amssymb}
\usepackage{latexsym}
\usepackage{multirow}
\usepackage{epsfig}
\usepackage{graphics}
\usepackage{graphicx}
\usepackage{mathrsfs}
\usepackage{subfigure}
\usepackage{mathrsfs}
\usepackage{bbding}
\usepackage{cite}
\usepackage{color}
\usepackage{xcolor}
\usepackage{booktabs}
\usepackage{amssymb,mathrsfs,amsmath}

\usepackage{flushend}
\usepackage{stfloats}
\usepackage{hyperref}
\usepackage{amsthm}
\theoremstyle{plain}
\newtheorem{rem}{Remark}

\allowdisplaybreaks [4]

\begin{document}	
	
%\title{Joint Deployment and Beamforming Design for Cost-Effective Coverage in IRS-Aided Movable Antenna Systems}
\title{Two-Scale Spatial Deployment for Cost-Effective Wireless Networks via Cooperative IRSs and Movable Antennas}
\author{Ying~Gao, Qingqing~Wu, Ziyuan~Zheng, Yanze~Zhu, Wen~Chen, Xin~Lin, and Shanpu Shen \vspace{-2mm} % <-this % stops a space	
\thanks{Y.~Gao, Q.~Wu, Z.~Zheng, Y.~Zhu, and W.~Chen are with the Department of Electronic Engineering, Shanghai Jiao Tong University, Shanghai 201210, China (e-mail: yinggao@sjtu.edu.cn; qingqingwu@sjtu.edu.cn; zhengziyuan2024@sjtu.edu.cn; yanzezhu@sjtu.edu.cn; wenchen@sjtu.edu.cn). X.~Lin is with the Shanghai Institute of Satellite Engineering, Shanghai 201109, China (e-mail: scar07@sina.com). S.~Shen is with the State Key Laboratory of Internet of Things for Smart City and the Department of Electrical and Computer Engineering, University of Macau, Macau, China (e-mail: shanpushen@um.edu.mo).}}% <-this % stops a space	

\maketitle

\begin{abstract}
	This paper proposes a two-scale spatial deployment strategy to ensure reliable coverage for multiple target areas, integrating macroscopic intelligent reflecting surfaces (IRSs) and fine-grained movable antennas (MAs). Specifically, IRSs are selectively deployed from candidate sites to shape the propagation geometry, while MAs are locally repositioned among discretized locations to exploit small-scale channel variations. The objective is to minimize the total deployment cost of MAs and IRSs by jointly optimizing the IRS site selection, MA positions, transmit precoding, and IRS phase shifts, subject to the signal-to-noise ratio (SNR) requirements for all target areas. This leads to a challenging mixed-integer non-convex optimization problem that is intractable to solve directly. To address this, we first formulate an auxiliary problem to verify the feasibility. A penalty-based double-loop algorithm integrating alternating optimization and successive convex approximation (SCA) is developed to solve this feasibility issue, which is subsequently adapted to obtain a suboptimal solution for the original cost minimization problem. Finally, based on the obtained solution, we formulate an element refinement problem to further reduce the deployment cost, which is solved by a penalty-based SCA algorithm. Simulation results demonstrate that the proposed designs consistently outperform benchmarks relying on independent area planning or full IRS deployment in terms of cost-efficiency. Moreover, for cost minimization, MA architectures are preferable in large placement apertures, whereas fully populated FPA architectures excel in compact ones; for worst-case SNR maximization, MA architectures exhibit a lower cost threshold for feasibility, while FPA architectures can attain peak SNR at a lower total cost.  %\looseness=-1
\end{abstract}

\begin{IEEEkeywords}
	Intelligent reflecting surface, movable antenna, two-scale spatial deployment, cost minimization, beamforming optimization. %\looseness=-1
\end{IEEEkeywords}

\vspace{-3mm}
\section{Introduction}
Multiple-input multiple-output (MIMO) technology serves as a cornerstone of modern wireless networks, leveraging spatial diversity to optimize spectral efficiency \cite{2004_Paulraj_overviewMIMO}. As the industry transitions from the fifth-generation (5G) to the sixth-generation (6G) era, the severe shortage of available spectrum has necessitated the deployment of ultra-massive antenna arrays at base stations (BSs) to unlock high degrees of spatial multiplexing \cite{2014_Larsson_massiveMIMO}. However, while these large-scale systems offer unprecedented connectivity, they impose significant burdens in terms of hardware expenditure and operational energy consumption. Consequently, developing solutions that balance superior performance with cost-effectiveness and low complexity has become a critical research priority \cite{2020_Chowdhury_challenges}.

Intelligent reflecting surface (IRS) has emerged as a pivotal enabler for green and cost-effective communications \cite{2019_Qingqing_Joint}. By manipulating incident waves via massive passive elements, the IRS proactively reconfigures the radio environment to suppress interference and enhance link reliability. Beyond their lightweight structure and flexible deployment capabilities, these devices offer a squared power gain that scales with the number of elements \cite{2019_Qingqing_Joint,2025_Qingqing_deployment}. This significant potential has sparked extensive research into integrating IRSs with modern wireless networks \cite{2020_Qingqing_Discrete,2019_Miao_Secure,2020_Xinrong_AN,2022_Weidong_IRS,2023_Ying_SWIPT,2025_Qiaoyan_Rotatable,2025_Qiaoyan_Cooperative}.
In particular, numerous studies have been devoted to optimizing IRS-assisted networks for eliminating coverage blind spots \cite{2021_Haiquan_coverage,2023_Guangji_coverage,2023_Jie_coverage_dualIRS,2024_Hanan_coverage,2023_Weidong_Coverage,2025_Min_Coverage}. For instance, the authors of \cite{2021_Haiquan_coverage} considered the co-design of active and passive beamforming alongside the placement of an aerial IRS to improve the minimum signal-to-noise ratio (SNR) in a target area. In a different context, the work in \cite{2023_Guangji_coverage} studied a static regulated IRS with distributed MIMO technology. By coordinating multiple access points, this architecture effectively utilizes spatial diversity to enhance wireless coverage. Additionally, the authors of \cite{2025_Min_Coverage} investigated the cost-effective deployment of multiple IRSs by jointly optimizing the site selection and physical parameters to satisfy a specific coverage rate requirement. 

Despite the channel reconfiguration capabilities of IRSs, the transceiver architectures are traditionally constrained by fixed-position antenna (FPA) geometries. Due to the lack of flexibility, such static designs limit the exploitation of spatial diversity and remain vulnerable to deep fading \cite{2023_Lipeng_overview}. To overcome this limitation, movable antenna (MA) technology \cite{2023_Lipeng_Modeling}, also known as fluid antenna \cite{2021_Wong_fluid}, has been introduced to enable flexible mechanical position adjustment within a confined region. By actively reconstructing the local channel environment to harvest the spatial variation gain fully, MAs achieve superior spectral efficiency and interference mitigation with fewer radio frequency (RF) chains \cite{2023_Lipeng_Modeling,2023_Wenyan_MIMO,2023_Lipeng_uplink}. Motivated by these advantages, extensive research efforts have been devoted to applying MAs to diverse domains, including multiuser networks \cite{2024_Haoran_MA,2024_Zhenyu_uplink,2024_Ziyuan_twotime}, interference channels \cite{2024_Honghao_IFC}, multicast transmission \cite{2024_Ying_multicast}, physical layer security \cite{2024_Guojie_secure_lett}, cognitive radio \cite{2024_Weidong_MA_spectrumsharing}, and wireless-powered communications \cite{2025_Ying_WPCN}. \looseness=-1

While both IRS and MA technologies are promising candidates for next-generation networks, they operate on fundamentally different principles. An IRS serves as a passive array that reshapes the propagation environment through phase adjustments and requires an external source for signal transmission. Conversely, an MA actively optimizes channel quality by mechanically maneuvering the antenna position within a confined region. These distinct characteristics create a natural synergy between the two paradigms. Specifically, the IRS is effective at establishing virtual links to illuminate coverage blind spots, while the MA excels at harvesting spatial diversity across the continuous field. Furthermore, the deployment of IRSs enriches the multipath components in the environment, and this enables MAs to exploit spatial channel variations more efficiently \cite{2025_Lipeng_Tutorial}. Recent studies confirm that integrating IRSs with MAs yields superior performance compared to conventional IRS-aided FPA systems \cite{2025_Qingqing_MAIRS}. Significant improvements have been reported in terms of outage probability \cite{2024_Rostami_fluidRIS}, data throughput \cite{2025_Yunan_MARIS,2024_Junteng_fluidRIS,2025_Weidong_MAIRS}, physical layer security \cite{2025_Rostami_fluidRIS_secrecy}, and coverage performance \cite{2025_Ying_Coverage}.

Despite the potential revealed in prior studies, two fundamental limitations remain. First, existing research on IRS-aided MA systems typically assumes fixed IRS deployment locations. This overlooks the practical need for optimal site selection in real network planning. Second, regarding the dimensioning of active and passive hardware, the authors of \cite{2025_Ying_Coverage} provided a preliminary analysis of the trade-off between the number of MAs and IRS elements. However, that study primarily investigated the worst-case SNR performance under a fixed total budget by evaluating discrete combinations of antenna and element counts. Such a simulation-based performance-maximization approach is inapplicable to the practical network planning problem where the goal is to satisfy strict quality-of-service (QoS) requirements with minimal expenditure. Since the optimal hardware scale is unknown a priori and coupled with deployment positions, relying on the exhaustive search of discrete configurations becomes computationally prohibitive and inefficient. Consequently, there is a lack of systematic strategies for cost-effective network planning that jointly optimize the deployment of hardware resources and the configuration of transmission parameters to guarantee service quality. \looseness=-1
 
\begin{figure}[!t]
	\vspace{-1mm}
	\centering
	\includegraphics[width = 0.49\textwidth]{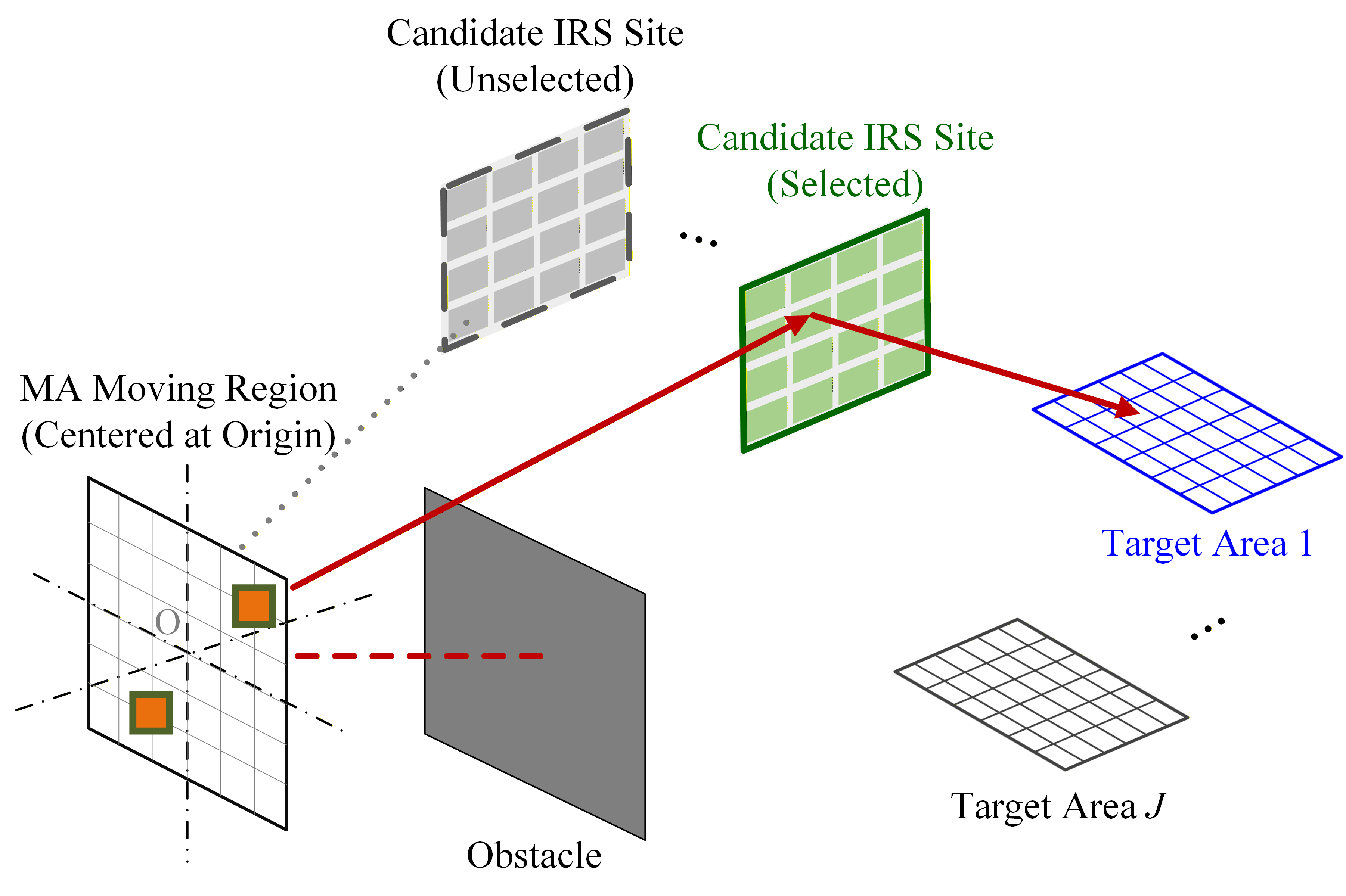}
	\caption{Illustration of an IRS-aided MA system, involving joint IRS and MA deployment and configuration for cost-effective coverage.} \label{Fig:system_model}
	\vspace{-3mm}
\end{figure}

Motivated by the above discussions, we study the cost-effective coverage problem in an IRS-aided MA system, as shown in Fig. \ref{Fig:system_model}. This system comprises a BS with a reconfigurable number of MAs and a set of candidate locations for IRS deployment, serving multiple target areas blocked by obstacles. Our main contributions are summarized as follows:
\begin{itemize}
	\item To the best of the authors' knowledge, this is the first work to study two-scale spatial deployment for cost-effective IRS-aided MA systems. We establish a novel optimization framework where macroscopic IRS site selection shapes propagation geometry and microscopic MA placement exploits local channel variations. Specifically, a cost minimization problem is formulated by jointly optimizing these spatial parameters together with transmit precoding and IRS phase shifts, subject to the SNR requirements of all target areas. The resulting formulation involves binary deployment decisions and non-convex variable coupling, leading to a mixed-integer non-convex program that is challenging to solve directly. %\looseness=-1
	\item We develop a hierarchical optimization framework to tackle the mixed-integer non-convexity and coupling hurdles. Specifically, a penalty-based double-loop algorithm integrating alternating optimization (AO) and successive convex approximation (SCA) is first designed to solve an auxiliary feasibility problem, ensuring a reliable initialization. This framework is then adapted for cost minimization, where the complex trilinear variable coupling is resolved via tight convex-hull linearization. Finally, an element-level refinement stage is introduced to prune hardware redundancy using a penalty-based SCA approach, further enhancing cost-efficiency while maintaining SNR guarantees. %\looseness=-1
	\item Numerical results show that: 1) A smaller MA step size improves feasibility with respect to the SNR constraints only when the resulting grid discretization preserves a non-decreasing maximum deployable antenna count $M_{\max}$ under the minimum inter-MA distance constraint. 2) Compared with benchmarks based on independent area planning or full IRS activation, the proposed designs consistently achieve lower deployment costs. 3) Regarding cost minimization, MAs are superior in large placement apertures while FPAs with $M_{\max}$ antennas suit compact ones; conversely, for worst-case SNR maximization, MAs offer a lower feasibility threshold, whereas such FPAs achieve peak performance at a lower total cost. %\looseness=-1 
\end{itemize}

The remainder of this paper is organized as follows. Section \ref{sec:model_and_formulation} presents the system model and formulates the deployment cost minimization problem. Before solving the problem, Section~\ref{sec:feasib_check} introduces a method to verify its feasibility. Section~\ref{sec:P1_solution} then proposes a computationally efficient algorithm to obtain a suboptimal solution of the problem. Building upon this, Section \ref{sec:element_sizing} refines the number of IRS elements to further reduce the deployment cost. Numerical results are presented in Section \ref{sec:simulation}, followed by the conclusion in Section \ref{sec:_conclusion}.

\emph{Notations:} Let $\mathbb{C}$ and $\mathbb{C}^{M \times N}$ denote the set of complex numbers and the space of $M \times N$ complex matrices, respectively. The superscripts $(\cdot)^T$ and $(\cdot)^H$ represent transpose and conjugate transpose, respectively. For any vector $\mathbf{x}$, $\|\mathbf{x}\|$ and $[\mathbf{x}]_i$ denote its Euclidean norm and $i$-th entry, respectively. Similarly, $[\mathbf{X}]_{i,j}$ denotes the $(i,j)$-th entry of matrix $\mathbf{X}$, and $\mathbf X(:,m)$ denotes its $m$-th column. The operators ${\rm diag}(\cdot)$ and ${\rm blkdiag}(\cdot)$ construct diagonal and block-diagonal matrices, respectively, while ${\rm Diag}(\cdot)$ extracts the main diagonal of a matrix as a vector. For a scalar $x\in\mathbb{C}$, $|x|$ and $\Re\{x\}$ denote its magnitude and real component, with $\jmath^2=-1$. $\mathbf{A}\succeq\mathbf{B}$ means $\mathbf{A}-\mathbf{B}$ is positive semidefinite. Additionally, $|\mathcal{K}|$ is the size of set $\mathcal{K}$, and $\mathcal{CN}(\boldsymbol{\mu},\mathbf{\Sigma})$ denotes the circularly symmetric complex Gaussian distribution with mean vector $\boldsymbol{\mu}$ and covariance matrix $\mathbf{\Sigma}$. $\odot$ represents the Hadamard product. 

%\vspace{-2mm}
\section{System Model and Problem Formulation}\label{sec:model_and_formulation}
\subsection{System Model}
As shown in Fig. \ref{Fig:system_model}, we consider a downlink system providing area-wide coverage over $J$ spatially separated target areas, indexed by $\mathcal J \triangleq \left\lbrace1,\cdots, J\right\rbrace$. A BS is deployed at a remote location due to practical site planning constraints, such as site availability, land-lease cost, urban planning regulations, and backhaul access. Furthermore, severe terrestrial blockage obstructs the direct paths between the BS and target areas, rendering these links negligible. To address this connectivity issue, multiple passive IRSs are deployed at elevated positions to create virtual line-of-sight (LoS) links. We assume that there are $L$ candidate sites available for IRS deployment, indexed by $\mathcal L \triangleq \left\lbrace1,\cdots,L\right\rbrace $. 

The BS is equipped with MAs connected to RF chains via flexible cables, enabling dynamic positioning within a square region $\mathcal C$ of size $A\times A$. We adopt a three-dimensional (3D) Cartesian coordinate system with the BS reference point at the origin, where $\mathcal C$ lies in the $y$–$z$ plane. Due to the fixed step size of electromechanical steppers, the antenna movement is quantized. The feasible MA locations are modeled as a discrete set of $M$ grid points with coordinates $\mathbf t_m \in \mathbb R^{3\times 1}$ for $m \in \mathcal M \triangleq \left\lbrace1,\cdots,M \right\rbrace$, arranged with a uniform step size of $d$. In the same coordinate system, a deployed IRS at candidate site $\ell\in\mathcal L$ has element positions $\left\lbrace \mathbf p_{\ell,n}\right\rbrace_{n=1}^{N_\ell} \subset \mathbb R^{3\times 1}$, with $\mathbf p_{\ell,1}$ serving as the reference point. Furthermore, let $\mathcal A_j \subset \mathbb R^{3\times 1}$ denote the $j$-th target area and $\mathbf u_j\in\mathcal A_j$ denote an arbitrary location in that area. 

The $J$ target areas are served sequentially, following either a round-robin cycle or a predetermined schedule. Accordingly, we define a binary variable $x_{m,j}\in\{0,1\}$ to indicate whether candidate location $m$ is occupied by an MA when serving area $j$ ($x_{m,j}=1$ if occupied and $0$ otherwise). The MA configuration is area-adaptive, i.e., $x_{m,j}$ may differ from $x_{m,j'}$ for $j\neq j'$. To enforce a minimum inter-MA separation, we impose the following constraint:
\begin{align}
	x_{m,j} + x_{q,j} \leq 1, \ \forall (m,q) \in \mathcal D, j\in\mathcal J,
\end{align}
where the distance-based conflict set $\mathcal D$ is defined as $\mathcal D \triangleq \left\lbrace (m,q)| m < q, \|\mathbf t_m - \mathbf t_q\| <  D \right\rbrace$. This constraint ensures that no two MAs are simultaneously deployed at locations whose distance is smaller than $D$ meters (m). In parallel, let $z_\ell\in\{0,1\}$ indicate whether an IRS is installed at candidate site $\ell$, where $z_\ell=1$ means selected and $0$ otherwise. While $z_\ell$ is area-independent, the IRS phase shifts are area-adaptive. For a deployed IRS at site $\ell$, the diagonal phase-shift matrix for area $j$ is given by $\mathbf \Theta_{\ell,j} = {\rm diag}\left(e^{\jmath \psi_{\ell,1,j}}, \ldots, e^{\jmath \psi_{\ell,N_\ell,j}}\right) \in \mathbb C^{N_\ell\times N_\ell}$, with $\psi_{\ell,n,j} \in [0,2\pi)$ denoting the phase shift of element $n \in \mathcal N_{\ell} \triangleq \left\lbrace 1, \ldots, N_\ell\right\rbrace$. 

The MA moving region and the physical size of each IRS are assumed to be much smaller than the corresponding link distances. Hence, all propagation paths can be treated as operating in the far field. Under this condition, relocating an MA influences only the phase of the associated channel coefficient, whereas the angle of departure (AoD), angle of arrival (AoA), and amplitude can be regarded as invariant \cite{2023_Lipeng_Modeling}. Furthermore, IRS-related links are treated as LoS-dominant under the assumption that the candidate sites are deliberately selected at elevated, unobstructed locations. If an IRS is placed at candidate location $\ell\in\mathcal L$ (i.e., $z_\ell = 1$), the BS-to-IRS channel when serving target area $j$ can be modeled as
\begin{align}\label{eq:channel_BS_IRS}
    & \hspace{1mm}\mathbf G_{\ell}(\mathbf x_j) = \sqrt{C_0d_{\rm \ell}^{-2}} \nonumber\\
	& \times \!\left[1, e^{\jmath\frac{2\pi}{\lambda}\left( \mathbf p_{\ell,2}-\mathbf p_{\ell,1}\right)^T\mathbf a(\theta^{\rm r}_\ell, \phi^{\rm r}_\ell)}, \ldots, e^{\jmath\frac{2\pi}{\lambda}\left( \mathbf p_{\ell,N_\ell} - \mathbf p_{\ell,1}\right)^T\mathbf a(\theta^{\rm r}_\ell, \phi^{\rm r}_\ell)}\!\right]^H \nonumber\\
	& \times \left(\mathbf x_j^T \odot \left[e^{\jmath\frac{2\pi}{\lambda}\mathbf t_1^T\mathbf a(\theta^{\rm t}_\ell, \phi^{\rm t}_\ell)}, \ldots, e^{\jmath\frac{2\pi}{\lambda}\mathbf t_M^T\mathbf a(\theta^{\rm t}_\ell, \phi^{\rm t}_\ell)}\right]\right)  \in \mathbb C^{N_\ell\times M},
\end{align}
with $\mathbf x_j \triangleq [x_{1,j},\ldots,x_{M,j}]^{T}\in\{0,1\}^{M}$ representing the MA configuration for area $j$, and $d_\ell \triangleq \|\mathbf p_{\ell,1}\|$ the distance from the BS reference point to the IRS reference point. The scalar $C_0 d_{\ell}^{-2}$ accounts for the large-scale path loss, with $C_0$ denoting the path-loss coefficient at the reference distance of 1 m. In addition, $\mathbf a(\theta^{\rm t}_\ell, \phi^{\rm t}_\ell) \triangleq \left[\cos\theta^{\rm t}_\ell\cos\phi^{\rm t}_\ell,\cos\theta^{\rm t}_\ell\sin\phi^{\rm t}_\ell, \sin\theta^{\rm t}_\ell\right]^T$ and $\mathbf a(\theta^{\rm r}_\ell, \phi^{\rm r}_\ell) \triangleq \left[\cos\theta^{\rm r}_\ell\cos\phi^{\rm r}_\ell, \cos\theta^{\rm r}_\ell\sin\phi^{\rm r}_\ell, \sin\theta^{\rm r}_\ell\right]^T$ represent the normalized transmit and receive direction vectors, respectively \cite{2023_Lipeng_uplink}. The parameters $\theta^{\rm t}_\ell$ and $\phi^{\rm t}_\ell$ are the elevation AoD and azimuth AoD at the BS, while $\theta^{\rm r}_\ell$ and $\phi^{\rm r}_\ell$ are the elevation AoA and azimuth AoA at the IRS. To express $\mathbf G_{\ell}(\mathbf x_j)$ in a more compact form, we further define $\mathbf e_\ell\triangleq \left[1, e^{\jmath\frac{2\pi}{\lambda}\left(\mathbf p_{\ell,2}-\mathbf p_{\ell,1}\right)^T\mathbf a(\theta^{\rm r}_\ell, \phi^{\rm r}_\ell)}, \ldots, e^{\jmath\frac{2\pi}{\lambda}\left( \mathbf p_{\ell,N_\ell} - \mathbf p_{\ell,1}\right)^T\mathbf a(\theta^{\rm r}_\ell, \phi^{\rm r}_\ell)}\right]^H \in\mathbb C^{N_{\ell}\times 1}$, $\mathbf g^H_\ell\triangleq  \left[e^{\jmath\frac{2\pi}{\lambda}\mathbf t_1^T\mathbf a(\theta^{\rm t}_\ell, \phi^{\rm t}_\ell)}, \ldots, e^{\jmath\frac{2\pi}{\lambda}\mathbf t_M^T\mathbf a(\theta^{\rm t}_\ell, \phi^{\rm t}_\ell)}\right] \in \mathbb C^{1\times M}$, $\mathbf X_j \triangleq \mathrm{diag}(\mathbf x_j)$, and $\overline{\mathbf G}_\ell \triangleq \sqrt{C_0d_{\rm \ell}^{-2}}\mathbf e_\ell\mathbf g^H_\ell$. Then, $\mathbf G_{\ell}(\mathbf x_j)$ can be rewritten as 
\begin{align}
	\mathbf G_{\ell}(\mathbf x_j) & = \sqrt{C_0d_{\rm \ell}^{-2}}\mathbf e_\ell\left(\mathbf x_j^T \odot \mathbf g^H_\ell\right) = \sqrt{C_0d_{\rm \ell}^{-2}}\mathbf e_\ell\mathbf g^H_\ell\mathbf X_j \nonumber\\
	& =\overline{\mathbf G}_\ell\mathbf X_j. 
\end{align}
Similarly, the channel from the IRS at candidate site $\ell$ to location $\mathbf u_j$ is modeled as
\begin{align}
	& \mathbf h_\ell^H(\mathbf u_j) = \sqrt{C_0\tilde d_{\ell}(\mathbf u_j)^{-2}} \nonumber\\
	& \times \left[1, e^{\jmath\frac{2\pi}{\lambda}\left(\mathbf p_{\ell,2} -\mathbf p_{\ell,1}\right)^T\mathbf a\left(\theta^{\rm t}_{\ell,j}(\mathbf u_j),\phi^{\rm t}_{\ell,j}(\mathbf u_j)\right)}, \ldots,\right.  \nonumber\\
	& \left. \quad e^{\jmath\frac{2\pi}{\lambda}\left(\mathbf p_{\ell,N_\ell}-\mathbf p_{\ell,1}\right) ^T\mathbf a\left(\theta^{\rm t}_{\ell,j}(\mathbf u_j),\phi^{\rm t}_{\ell,j}(\mathbf u_j)\right)}\right]  \in \mathbb C^{1\times N_\ell},
\end{align} 
where $\tilde d_{\ell}(\mathbf u_j) \triangleq \left\|\mathbf p_{\ell,1} - \mathbf u_j\right\|$ denotes the distance between the IRS reference point and $\mathbf u_j$, and $\mathbf a\left(\theta^{\rm t}_{\ell,j}(\mathbf u_j),\phi^{\rm t}_{\ell,j}(\mathbf u_j)\right) \triangleq \Big[\!\cos\theta^{\rm t}_{\ell,j}\!(\mathbf u_j)\!\cos\phi^{\rm t}_{\ell,j}\!(\mathbf u_j),\cos\theta^{\rm t}_{\ell,j}\!(\mathbf u_j)\!\sin\phi^{\rm t}_{\ell,j}\!(\mathbf u_j),\sin\theta^{\rm t}_{\ell,j}\!(\mathbf u_j)\!\Big]^T$ denotes the normalized wave vector, with $\theta^{\rm t}_{\ell,j}(\mathbf u_j)$ and $\phi^{\rm t}_{\ell,j}(\mathbf u_j)$ being the elevation and azimuth AoDs towards $\mathbf u_j$, respectively. 

Let $P$ denote the transmit power and define an $M$-dimensional beamforming vector $\mathbf w_{\mathbf u_j} \in \mathbb C^{M\times 1}$ associated with all candidate MA positions for location $\mathbf u_j$. The effective beamforming vector is then given by 
\begin{align}
	\tilde{\mathbf{w}}_{\mathbf u_j} = \mathbf X_j\mathbf{w}_{\mathbf u_j} \in \mathbb C^{M\times 1},
\end{align}
where $\tilde{\mathbf w}_{\mathbf u_j}$ retains the beamforming coefficients at candidate positions occupied by MAs and forces the coefficients at all other positions to zero. Neglecting the signal components resulting from double or multiple reflections at the deployed IRSs, the received SNR at location $\mathbf u_j\in\mathcal A_j$, $j\in\mathcal J$ can be written as 
\begin{align}\label{eq:SNR}
    \gamma_{\mathbf u_j} & = \bar P\left|\left(\sum_{\ell=1}^Lz_\ell\mathbf h_{\ell}^H(\mathbf u_j)\mathbf \Theta_{\ell,j}\overline{\mathbf G}_\ell\mathbf X_j\right)\mathbf X_j\mathbf w_{\mathbf u_j}\right|^2 \nonumber\\
	& = \bar P\left|\left(\sum_{\ell=1}^Lz_\ell\mathbf h_{\ell}^H(\mathbf u_j)\mathbf \Theta_{\ell,j}\overline{\mathbf G}_\ell\right)\mathbf X_j\mathbf w_{\mathbf u_j}\right|^2, 
\end{align}
where $\bar P \triangleq \frac{P}{\sigma^2}$ with $\sigma^2$ representing the noise power at location $\mathbf u$, and the second equality follows from the idempotence of the binary diagonal matrix $\mathbf X_j$, i.e., $\mathbf X_j^2 = \mathbf X_j$. 

Let $c_{\rm MA}$ denote the unit cost per MA. Since the $J$ target areas are served sequentially and the same hardware can be reused across areas, the required number of MAs is determined by the maximum number of active MAs over all areas, i.e., $\max_{j\in\mathcal J}\left\lbrace\sum_{m=1}^Mx_{m,j}\right\rbrace$. The corresponding MA deployment cost is therefore modeled as $c_{\rm MA}\max_{j\in\mathcal J}\left\lbrace \sum_{m=1}^Mx_{m,j}\right\rbrace$. Similarly, the IRS deployment cost is given by $\sum_{\ell=1}^L z_\ell\left( c_\ell +  c_{\rm e} N_\ell\right) $, where $c_\ell$ is the site-specific installation cost at candidate site $\ell$ (reflecting, e.g., terrain, leasing and installation complexity), and $c_{\rm e}$ denotes the unit cost per IRS element. Combining both contributions, the total infrastructure deployment cost is 
\begin{align}
	C_{\rm tot} = c_{\rm MA}\max_{j\in\mathcal J}\left\lbrace \sum_{m=1}^Mx_{m,j}\right\rbrace + \sum_{\ell=1}^L z_\ell\left( c_\ell +  c_{\rm e} N_\ell\right). 
\end{align}

\subsection{Problem Formulation}
In this paper, we aim to minimize the total deployment cost by jointly selecting the IRS sites, determining the MA placements and hence the number of deployed MAs, optimizing the IRS phase shifts and the BS transmit beamforming vectors, while guaranteeing SNR requirements for all target areas. The problem of interest can be formulated as 
\begin{subequations}\label{P1}
	\begin{eqnarray}
		& \text{(P1)}:& \underset{\substack{\left\lbrace x_{m,j}\right\rbrace,\left\lbrace z_\ell\right\rbrace,\\\left\lbrace \mathbf \Theta_{\ell,j}\right\rbrace,\left\lbrace \mathbf w_{\mathbf u_j}\right\rbrace }}{\min} \hspace{2mm} c_{\rm MA}\max_{j\in\mathcal J}\left\lbrace \sum_{m=1}^Mx_{m,j}\right\rbrace \nonumber\\
		&& \hspace{2cm} + \sum_{\ell=1}^L z_\ell\left( c_\ell +  c_{\rm e} N_\ell\right) \\
		& \text{s.t.}& \hspace{-2mm} \bar P\left|\left(\sum_{\ell=1}^Lz_\ell\mathbf h_{\ell}^H(\mathbf u_j)\mathbf \Theta_{\ell,j}\overline{\mathbf G}_\ell\right)\mathbf X_j\mathbf w_{\mathbf u_j}\right|^2 \geq \gamma_{\rm th, \it j}, \nonumber\\
		&& \hspace{-2mm} \forall \mathbf u_j \in \mathcal A_j, j\in\mathcal J, \label{P1_cons:b}\\
		&& \hspace{-2mm} \left\|\mathbf X_j\mathbf w_{\mathbf u_j}\right\| = 1, \ \forall u_j \in\mathcal A_j, \label{P1_cons:c}\\
		&& \hspace{-2mm} \mathbf X_j = {\rm diag}\left(x_{1,j},x_{2,j},\ldots,x_{M,j}\right), \ \forall j\in\mathcal J, \label{P1_cons:d} \\
		&& \hspace{-2mm} x_{m,j} \in \{0,1\} , \ \forall m\in\mathcal M, j\in\mathcal J, \label{P1_cons:e}\\ 
		&& \hspace{-2mm} x_{m,j} + x_{q,j} \leq 1, \ \forall (m,q) \in \mathcal D, j\in\mathcal J, \label{P1_cons:f}\\
		&& \hspace{-2mm} \sum_{m=1}^M x_{m,j} \leq M_{\max}, \forall j\in\mathcal J, \label{P1_cons:g}\\
		&& \hspace{-2mm} z_\ell \in \{0,1\}, \ \forall \ell\in\mathcal L, \label{P1_cons:h}\\
		&& \hspace{-2mm} \sum_{\ell=1}^L z_\ell \leq L, \label{P1_cons:i}\\
		&& \hspace{-2mm} \left|\left[ \mathbf \Theta_{\ell,j}\right]_{n,n} \right| = 1, \ \forall  \ell\in\mathcal L, j\in\mathcal J, n\in\mathcal N_\ell, \label{P1_cons:j} 
	\end{eqnarray}
\end{subequations} 
where \eqref{P1_cons:b} ensures that the received SNR at any location $\mathbf u_j$ in area $j$ is no less than a predefined threshold $\gamma_{\rm th, \it j}$, and \eqref{P1_cons:c} ensures that the effective beamforming vector for $\mathbf u_j$ has an unit norm. For the MA configuration, \eqref{P1_cons:d} and \eqref{P1_cons:e} define the binary placement decisions. The selected MA positions are subject to the minimum-separation requirement captured by the conflict set $\mathcal D$ in \eqref{P1_cons:f}, which, together with the moving region and the grid step size $d$, specifies the upper bound $M_{\max}$ on the number of simultaneously deployable MAs in \eqref{P1_cons:g}. Finally, \eqref{P1_cons:h}-\eqref{P1_cons:j} limit the total number of deployable IRSs and imposes the modulus constraint on the IRS reflection coefficients. \looseness=-1

Problem (P1) is challenging to solve due to: 1) the hybrid discrete-continuous variables; 2) the semi-infinite SNR constraints arising from continuous target areas; and 3) the inherent non-convexity of the SNR and unit-modulus constraints. Hence, (P1) is a mixed-integer non-convex program, and computing a globally optimal solution is generally prohibitive. 

\section{Feasibility Check}\label{sec:feasib_check}
Before solving problem (P1), we first check its feasibility. Specifically, note that if the SNR targets in (P1) cannot be satisfied even under the most favorable deployment configuration, then problem (P1) is infeasible regardless of the cost coefficients. To this end, we consider an auxiliary feasibility check problem where all candidate IRSs are activated and the BS is permitted to deploy at most $M_{\max}$ MAs for each target area. Concretely, we set $z_\ell = 1$, $\forall \ell\in\mathcal L$, and introduce a non-negative slack variable $\eta$ as the common lower bound of the normalized SNR, namely, the received SNR divided by the corresponding threshold. The feasibility check problem is formulated as
\begin{subequations}\label{P2}
	\begin{eqnarray}
		& \text{(P2)}:& \underset{\substack{\left\lbrace x_{m,j}\right\rbrace,\left\lbrace \mathbf \Theta_{\ell,j}\right\rbrace, \\ \left\lbrace \mathbf w_{\mathbf u_j}\right\rbrace, \eta \geq 0}}{\max} \hspace{2mm} \eta\\
		&\text{s.t.}& \hspace{-2mm} \frac{\bar P}{\gamma_{\rm th, \it j}}\left|\left(\sum_{\ell=1}^L\mathbf h_{\ell}^H(\mathbf u_j)\mathbf \Theta_{\ell,j}\overline{\mathbf G}_\ell\right)\mathbf X_j\mathbf w_{\mathbf u_j}\right|^2 \geq \eta, \nonumber\\
		&& \hspace{-2mm} \forall \mathbf u_j \in \mathcal A_j, \ \forall j\in\mathcal J, \label{P2_cons:a}\\
		&& \hspace{-2mm} \eqref{P1_cons:c}-\eqref{P1_cons:g}, \eqref{P1_cons:j}.
	\end{eqnarray}
\end{subequations}
Note that the MA placement achieving the maximum cardinality $M_{\max}$ is not necessarily unique. Moreover, a fully populated array can be forced into geometrically suboptimal locations, and the SNR-optimal MA configuration may differ across target areas. Thus, even with $M_{\max}$ fixed, the placement variables $\left\lbrace x_{m,j}\right\rbrace$ must still be optimized. 
Let $\eta^*$ be the optimal value of problem (P2). If $\eta^* \geq 1$, the SNR requirements in problem (P1) are achievable, and we proceed to solve (P1) for cost minimization; otherwise, (P1) is infeasible. %\looseness=-1

Next, we solve the feasibility check problem (P2). It can be readily verified that the optimal transmit beamformer $\mathbf w_{\mathbf u_j}^*$ follows the maximum-ratio-transmission (MRT) principle: 
\begin{align}\label{P2_opt_w}
	\mathbf w_{\mathbf u_j}^* = \frac{\mathbf X_j\left(\sum_{\ell=1}^L\mathbf h_{\ell}^H(\mathbf u_j)\mathbf \Theta_{\ell,j}\overline{\mathbf G}_\ell\right)^H}{\left\| \mathbf X_j\left(\sum_{\ell=1}^L\mathbf h_{\ell}^H(\mathbf u_j)\mathbf \Theta_{\ell,j}\overline{\mathbf G}_\ell\right)^H\right\|}.
\end{align}
Substituting \eqref{P2_opt_w} into problem (P2) yields
\begin{subequations}\label{P2_eqv}
	\begin{eqnarray}
		&\hspace{-2mm}\underset{\left\lbrace x_{m,j}\right\rbrace,\left\lbrace \mathbf \Theta_{\ell,j}\right\rbrace, \eta \geq 0}{\max}& \eta\\
		&\text{s.t.}& \hspace{-8mm} \frac{\bar P}{\gamma_{\rm th, \it j}}\left\|\left(\sum_{\ell=1}^L\mathbf h_{\ell}^H(\mathbf u_j)\mathbf \Theta_{\ell,j}\overline{\mathbf G}_\ell\right)\mathbf X_j\right\|^2 \geq \eta, \nonumber\\
		&& \hspace{-8mm} \forall \mathbf u_j \in \mathcal A_j, \ \forall j\in\mathcal J, \label{P2_eqv_cons:a}\\
		&& \hspace{-8mm} \eqref{P1_cons:d}-\eqref{P1_cons:g}, \eqref{P1_cons:j}.
	\end{eqnarray}
\end{subequations}
Problem \eqref{P2_eqv} poses significant challenges due to the semi-infinite SNR constraints induced by the continuous target areas $\left\lbrace \mathcal A_j\right\rbrace$, the coupling of optimization variables in \eqref{P2_eqv_cons:a}, and the non-convex constraints \eqref{P1_cons:e} and \eqref{P1_cons:j}. To make the problem tractable, we first approximate the semi-infinite constraints by discretizing each $\mathcal A_j$ into a dense grid $\mathcal G_j$ and enforcing \eqref{P2_eqv_cons:a} only over $\mathcal G_j$, yielding %\looseness=-1
\begin{subequations}\label{P2_eqv_slack}
	\begin{eqnarray}
		&\hspace{-2mm}\underset{\left\lbrace x_{m,j}\right\rbrace,\left\lbrace \mathbf \Theta_{\ell,j}\right\rbrace, \eta \geq 0}{\max}& \eta\\
		&\text{s.t.}& \hspace{-8mm} \frac{\bar P}{\gamma_{\rm th, \it j}}\left\|\left(\sum_{\ell=1}^L\mathbf h_{\ell}^H(\mathbf u_j)\mathbf \Theta_{\ell,j}\overline{\mathbf G}_\ell\right)\mathbf X_j\right\|^2 \geq \eta, \nonumber\\
		&& \hspace{-8mm} \forall \mathbf u_j \in \mathcal G_j, \ \forall j\in\mathcal J, \label{P2_eqv_slack_cons:a}\\
		&& \hspace{-8mm} \eqref{P1_cons:d}-\eqref{P1_cons:g}, \eqref{P1_cons:j}.
	\end{eqnarray}
\end{subequations}
Problem \eqref{P2_eqv_slack} serves as a discretized approximation of problem \eqref{P2_eqv}. For static LoS channels, a high‑resolution offline grid can keep the approximation error small.
Although discretization removes infinitely many constraints, the variable coupling and non‑convexity persist. To address these remaining issues, we employ an AO framework that alternately optimizes $\left\lbrace x_{m,j}\right\rbrace$ and $\left\lbrace \mathbf \Theta_{\ell,j}\right\rbrace$ until convergence, as detailed below. %\looseness=-1

%\hspace{-2mm}
\subsection{Optimizing $\left\lbrace x_{m,j}\right\rbrace$ for Given $\left\lbrace \mathbf \Theta_{\ell,j}\right\rbrace$} \label{sec:feasib_check_A}
When $\left\lbrace \mathbf \Theta_{\ell,j}\right\rbrace$ is fixed, problem \eqref{P2_eqv_slack} reduces to
\begin{eqnarray}\label{P2_eqv_slack2_sub1}
	\underset{\left\lbrace x_{m,j}\right\rbrace, \eta \geq 0}{\max} \hspace{2mm} \eta \hspace{8mm} \text{s.t.} \hspace{2mm}  \eqref{P2_eqv_slack_cons:a},\eqref{P1_cons:d}-\eqref{P1_cons:g}. 
\end{eqnarray}	
To facilitate the solution design, we expand the quadratic term $\left\|\left(\sum_{\ell=1}^L\mathbf h_{\ell}^H(\mathbf u_j)\mathbf \Theta_{\ell,j}\overline{\mathbf G}_\ell\right)\mathbf X_j\right\|^2$ in constraint \eqref{P2_eqv_slack_cons:a} as 
\begin{align}\label{SNR_expand_1}
	& \left\|\left(\sum_{\ell=1}^L\mathbf h_{\ell}^H(\mathbf u_j)\mathbf \Theta_{\ell,j}\overline{\mathbf G}_\ell\right)\mathbf X_j\right\|^2 \nonumber\\
	& = \sum_{m=1}^Mx_{m,j}\left|\sum_{\ell=1}^L\mathbf h_{\ell}^H(\mathbf u_j)\mathbf \Theta_{\ell,j}\overline{\mathbf G}_\ell(:,m)\right|^2 \nonumber\\
	& \triangleq \sum_{m=1}^Mx_{m,j}C_{m,j}(\mathbf u_j),
\end{align} 
where $C_{m,j}(\mathbf u_j) \triangleq \left|\sum_{\ell=1}^L\mathbf h_{\ell}^H(\mathbf u_j)\mathbf \Theta_{\ell,j}\overline{\mathbf G}_\ell(:,m)\right|^2$. 
Substituting \eqref{SNR_expand_1} into problem \eqref{P2_eqv_slack2_sub1} yields the following equivalent formulation: 
\begin{subequations}\label{P2_eqv_slack2_sub1_eqv}
	\begin{eqnarray}
		&\hspace{-1cm}\underset{\left\lbrace x_{m,j}\right\rbrace, \eta \geq 0}{\max}& \eta\\
		&\hspace{-1.4cm}\text{s.t.}& \hspace{-9mm} \frac{\bar P}{\gamma_{\rm th, \it j}} \sum_{m=1}^Mx_{m,j}C_{m,j}(\mathbf u_j) \geq \eta, \forall \mathbf u_j \in \mathcal G_j, \forall j\in\mathcal J, \label{P2_eqv_slack_cons:a_eqv} \\
		&& \hspace{-9mm} \eqref{P1_cons:e}-\eqref{P1_cons:g},
	\end{eqnarray}
\end{subequations}
Problem \eqref{P2_eqv_slack2_sub1_eqv} is a mixed-integer linear program (MILP). Although it can be solved by off-the-shelf MILP solvers (e.g., MOSEK), calling an exact integer solver at each AO iteration becomes computationally prohibitive in large-scale settings due to the exponential worst-case complexity of integer programming. To obtain a scalable, polynomial-time update that can be embedded into the iterative framework, we adopt a penalty-based continuous reformulation of the binary constraint \eqref{P1_cons:e}. Specifically, \eqref{P1_cons:e} is equivalent to the intersection of the following two inequalities:
\begin{subequations}\label{P1_cons:e_eqv}
	\begin{align}
		& 0 \leq x_{m,j} \leq 1, \ \forall m\in\mathcal M, j \in\mathcal J, \label{P1_cons:e_eqv1}\\
		& x_{m,j} - x_{m,j}^2 \leq 0, \ \forall m\in\mathcal M, j \in\mathcal J. \label{P1_cons:e_eqv2}
	\end{align}
\end{subequations}
Here, \eqref{P1_cons:e_eqv1} defines a convex set, whereas \eqref{P1_cons:e_eqv2} is a non-convex constraint in the difference-of-convex form. To tackle the non-convexity in \eqref{P1_cons:e_eqv2}, we adopt the SCA technique by linearizing the quadratic term $x_{m,j}^2$ via the first-order Taylor expansion at a given local point $x_{m,j}^r$ in the $r$-th iteration. This yields the following global affine lower bound: 
\begin{align}\label{ineq:x_lower_bound}
	 x_{m,j}^2 & \geq - \left( x_{m,j}^r\right)^2 + 2x_{m,j}^rx_{m,j} \nonumber\\
	 & \triangleq \Pi^{\rm lb, \it r}(x_{m,j}), \ \forall m\in\mathcal M, j\in\mathcal J.
\end{align}
Substituting this lower bound into \eqref{P1_cons:e_eqv2} yields the convex surrogate constraint $x_{m,j}-\Pi^{\rm lb,\it r}(x_{m,j})\le 0, \forall m\in\mathcal M, j\in\mathcal J$. However, imposing this SCA-based surrogate as a hard constraint together with the box constraint \eqref{P1_cons:e_eqv1} may result in overly conservative updates, which can significantly shrink the feasible region and even lead to algorithm stagnation in practice. To maintain sufficient update flexibility while still promoting integrality, we instead penalize the violation of the surrogate constraint in the objective of problem~\eqref{P2_eqv_slack2_sub1_eqv}. This gives the following penalized formulation: 
\begin{subequations}\label{P2_eqv_slack2_sub1_eqv_pena}
	\begin{eqnarray}
		&\hspace{-1cm}\underset{\left\lbrace x_{m,j}\right\rbrace, \eta \geq 0}{\max}& \eta - \rho\sum_{m=1}^M\sum_{j=1}^J\left( x_{m,j} - \Pi^{\rm lb, \it r}(x_{m,j})\right) \\
		&\hspace{-1.4cm}\text{s.t.}& \hspace{-7mm} \eqref{P1_cons:e_eqv1}, \eqref{P2_eqv_slack_cons:a_eqv}, \eqref{P1_cons:f}-\eqref{P1_cons:g},
	\end{eqnarray}
\end{subequations}
where $\rho > 0$ is a penalty factor used to control the tightness of the relaxation. Problem \eqref{P2_eqv_slack2_sub1_eqv_pena} is a convex linear program, which can be efficiently solved using standard tools such as CVX. As the SCA iterations proceed, the linearization becomes tight, and $x_{m,j}-\Pi^{\rm lb,r}(x_{m,j})$ approaches $x_{m,j}-x_{m,j}^2=x_{m,j}(1-x_{m,j})$. Since this term is subtracted in the maximization objective, a sufficiently large $\rho$ strongly discourages fractional $x_{m,j}\in(0,1)$ and drives $x_{m,j}-\Pi^{\rm lb,r}(x_{m,j})$ towards zero. Together with the box constraint $0\le x_{m,j}\le 1$, this promotes $x_{m,j}(1-x_{m,j})\to 0$, thereby pushing $x_{m,j}$ towards $\{0,1\}$. 

\subsection{Optimizing $\left\lbrace \mathbf \Theta_{\ell,j}\right\rbrace$ for Given $\left\lbrace x_{m,j}\right\rbrace$} \label{sec:feasib_check_B}

With $\left\lbrace x_{m,j}\right\rbrace$ fixed, the optimization of $\left\lbrace \mathbf \Theta_{\ell,j}\right\rbrace$ reduces to solving problem \eqref{P2_eqv_slack} subject to constraints \eqref{P2_eqv_slack_cons:a} and \eqref{P1_cons:j}. To streamline the formulation and expose the curvature of the key term, we define the aggregate dimension $N\triangleq \sum_{\ell=1}^L N_\ell$, the stacked channel vector
$\mathbf h^H(\mathbf u_j)\triangleq \left[ \mathbf h_1^H(\mathbf u_j),\ldots,\mathbf h_L^H(\mathbf u_j)\right] \in\mathbb C^{1\times N}$, the stacked channel matrix
$\bar{\mathbf G}\triangleq \left[ \bar{\mathbf G}_1^T,\ldots,\bar{\mathbf G}_L^T\right]^T\in\mathbb C^{N\times M}$, and the cascaded channel matrix $\mathbf \Psi(\mathbf u_j)\triangleq \mathrm{diag}\!\left(\mathbf h^H\left( \mathbf u_j\right) \right) \bar{\mathbf G}\in\mathbb C^{N\times M}$. Further, let $\mathbf \Theta_j\triangleq \mathrm{blkdiag}(\mathbf \Theta_{1,j},\ldots,\mathbf \Theta_{L,j})\in\mathbb C^{N\times N}$ and $\mathbf v_j\triangleq \mathrm{diag}(\mathbf \Theta_j^H)\in\mathbb C^{N\times 1}$. Then, the quadratic term in \eqref{P2_eqv_slack_cons:a} can be recast as 
\begin{align}\label{quad_term_eqv}         
	& \left\|\left(\sum_{\ell=1}^L\mathbf h_{\ell}^H(\mathbf u_j)\mathbf \Theta_{\ell,j}\overline{\mathbf G}_\ell\right)\mathbf X_j\right\|^2 = \left\|\mathbf h^H(\mathbf u_j)\mathbf \Theta_j\bar{\mathbf G}\mathbf X_j\right\|^2 \nonumber\\
    & \hspace{2.5cm} = \left\|\mathbf v_j^H\mathbf \Psi(\mathbf u_j)\mathbf X_j\right\|^2 \triangleq \mathbf v_j^H\mathbf R(\mathbf u_j)\mathbf v_j,
\end{align}
where $\mathbf R(\mathbf u_j) \triangleq \mathbf \Psi(\mathbf u_j)\mathbf X_j\mathbf X_j^H\mathbf \Psi^H(\mathbf u_j)$. Consequently, the subproblem is equivalent to 
\begin{subequations}\label{P2_eqv_slack2_sub2}
	\begin{eqnarray}
		&\hspace{-5mm}\underset{\left\lbrace \mathbf v_j \right\rbrace, \eta}{\max} & \eta\\
		&\hspace{-5mm}\text{s.t.}& \hspace{-3mm} \frac{\bar P}{\gamma_{\rm th, \it j}} \mathbf v_j^H\mathbf R(\mathbf u_j)\mathbf v_j \geq \eta, \ \forall \mathbf u_j \in \mathcal G_j, \forall j\in\mathcal J, \label{P2_eqv_slack2_sub2_cons:b}\\
		&& \hspace{-3mm} \left|\left[\mathbf v_j\right]_n\right|=1,\ \forall j\in\mathcal J,\ n\in\mathcal N, \label{P2_eqv_slack2_sub2_cons:c}
	\end{eqnarray}
\end{subequations} 
where $\mathcal N$ denotes the set of all IRS elements. Since $\mathbf R(\mathbf u_j)\succeq \mathbf 0$, the term $\mathbf v_j^H\mathbf R(\mathbf u_j)\mathbf v_j$ is convex, rendering constraint \eqref{P2_eqv_slack2_sub2_cons:b} non-convex. To address this, we adopt the SCA framework and linearize $\mathbf v_j^H\mathbf R(\mathbf u_j)\mathbf v_j$ at a given local point $\mathbf v_j^r$ to obtain the following affine lower bound: 
\begin{align}\label{ineq:quad_lb}
	\mathbf v_j^H\mathbf R(\mathbf u_j)\mathbf v_j & \geq 2\Re\left\lbrace \left(\mathbf v_j^r\right)^H\mathbf R(\mathbf u_j)\mathbf v_j\right\rbrace - \left(\mathbf v_j^r\right)^H\mathbf R(\mathbf u_j)\mathbf v_j^r \nonumber\\
	& \triangleq \Xi^{\rm lb,\it r}(\mathbf v_j, \mathbf u_j). 
\end{align} 
Accordingly, the non-convex constraint \eqref{P2_eqv_slack2_sub2_cons:b} is approximated by the following convex constraint:
\begin{align}
	\frac{\bar P}{\gamma_{\rm th, \it j}} \Xi^{\rm lb,\it r}(\mathbf v_j, \mathbf u_j) \geq \eta, \ \forall \mathbf u_j \in \mathcal G_j, \forall j\in\mathcal J.  \label{P2_eqv_slack2_sub2_cons:b_sca}
\end{align}

Next, we handle the unit-modulus constraints in \eqref{P2_eqv_slack2_sub2_cons:c}, which can be written as the following pair of inequalities:
\begin{subequations}\label{P2_eqv_slack2_sub2_cons:c_eqv}
	\vspace{-3.5mm}
	\begin{align}
		1 \le \left|\left[\mathbf v_j\right]_n\right|^2,\ \forall j\in\mathcal J,\ n\in\mathcal N, \label{P2_eqv_slack2_sub2_cons:c_eqv1}\\
		\left|\left[\mathbf v_j\right]_n\right|^2 \le 1,\ \forall j\in\mathcal J,\ n\in\mathcal N. \label{P2_eqv_slack2_sub2_cons:c_eqv2}
	\end{align}
\end{subequations}
It is clear that \eqref{P2_eqv_slack2_sub2_cons:c_eqv2} is convex, whereas \eqref{P2_eqv_slack2_sub2_cons:c_eqv1} is non-convex.
To obtain a tractable inner approximation of \eqref{P2_eqv_slack2_sub2_cons:c_eqv1}, we linearize $\left|\left[\mathbf v_j\right]_n\right|^2$
at the local point $[\mathbf v_j^{r}]_n$ as
\begin{align}
	\left|\left[\mathbf v_j\right]_n\right|^2
	& \ge 2\Re\!\left\{\left([\mathbf v_j^{r}]_n\right)^*\left[\mathbf v_j\right]_n\right\}
	-\left|[\mathbf v_j^{r}]_n\right|^2 \nonumber\\
	& \triangleq \chi^{\rm lb,\it r}\!\left(\left[\mathbf v_j\right]_n\right).
\end{align}
Replacing $\left|\left[\mathbf v_j\right]_n\right|^2$ in \eqref{P2_eqv_slack2_sub2_cons:c_eqv1} by $\chi^{\rm lb,\it r}\!\left(\left[\mathbf v_j\right]_n\right)$ yields the SCA surrogate constraint $1\le \chi^{\rm lb,\it r}\!\left(\left[\mathbf v_j\right]_n\right),\ \forall j\in\mathcal J,\ n\in\mathcal N$.
Similar to the previous subsection, enforcing this surrogate constraint along with the unit-disc constraint \eqref{P2_eqv_slack2_sub2_cons:c_eqv2} may be overly restrictive and hinder iterative progress. To alleviate this issue, we adopt the penalty method. However, unlike the binary case in \eqref{P2_eqv_slack2_sub1_eqv_pena}, directly penalizing $1-\chi^{\rm lb,\it r}\!\left(\left[\mathbf v_j\right]_n\right)$ in the objective function is generally undesirable. Minimizing $1-\chi^{\rm lb,\it r}\!\left(\left[\mathbf v_j\right]_n\right)$ is equivalent to maximizing the affine lower bound $\chi^{\rm lb,\it r}\!\left(\left[\mathbf v_j\right]_n\right)$, which, under the unit-disc constraint \eqref{P2_eqv_slack2_sub2_cons:c_eqv2}, reduces to maximizing $\Re\!\left\{\left([\mathbf v_j^{r}]_n\right)^*\left[\mathbf v_j\right]_n\right\}$. This tends to phase-align $\left[\mathbf v_j\right]_n$ with $\left[\mathbf v_j^r\right]_n$, leading to overly conservative updates. Such conservatism may cause premature stagnation and impede the improvement of $\eta$ over iterations.

Instead, we introduce non-negative slack variables $\{\delta_{j,n}\}$ to relax the SCA surrogate constraint as
\begin{align}\label{P2_eqv_slack2_sub2_cons:c_eqv1_sca_slack}
	1-\delta_{j,n}\le \chi^{\rm lb,\it r}\!\left(\left[\mathbf v_j\right]_n\right),
	\ \forall j\in\mathcal J,\ n\in\mathcal N,
\end{align}
where $\delta_{j,n}\ge 0$ quantifies the violation of \eqref{P2_eqv_slack2_sub2_cons:c_eqv1} in the current iteration. We then penalize $\{\delta_{j,n}\}$ in the objective. This yields a one-sided (hinge-loss) penalty: the penalty is activated only when $1>\chi^{\rm lb,\it r}\!\left([\mathbf v_j]_n\right)$, whereas $\delta_{j,n}=0$ once the surrogate constraint is satisfied, thereby avoiding an unnecessary incentive to further increase $\chi^{\rm lb,\it r}\!\left([\mathbf v_j]_n\right)$. 
Accordingly, the phase-update subproblem in the $r$-th SCA iteration is formulated as
\begin{subequations}\label{P2_eqv_slack2_sub2_pen}
	\begin{eqnarray}
		&\hspace{-5mm}\underset{\substack{\left\lbrace \mathbf v_j \right\rbrace, \eta, \\ \left\lbrace \delta_{j,n} \geq 0 \right\rbrace}}{\max} 
		& \eta - \lambda \sum_{j\in\mathcal J}\sum_{n\in\mathcal N} \delta_{j,n} \\
		&\hspace{-5mm}\text{s.t.}& \hspace{-5mm} \eqref{P2_eqv_slack2_sub2_cons:b_sca},  \eqref{P2_eqv_slack2_sub2_cons:c_eqv2}, \eqref{P2_eqv_slack2_sub2_cons:c_eqv1_sca_slack}, 
	\end{eqnarray}
\end{subequations} 
where $\lambda>0$ is a penalty factor. At optimality, the slack variable satisfies $\delta_{j,n} = \left[1-\chi^{\rm lb, \it r}\!\left(\left[\mathbf v_j\right]_n\right) \right]_{+}$. With a properly chosen (or gradually increased) $\lambda$, the slack variables $\{\delta_{j,n}\}$ tend to vanish, thereby promoting $\left|\left[\mathbf v_j\right]_n\right|\to 1$. Problem \eqref{P2_eqv_slack2_sub2_pen} is a convex quadratically constrained quadratic program (QCQP) and can be efficiently solved using standard tools, e.g., CVX.

\subsection{Overall Algorithm}
The proposed algorithm adopts a \textbf{double-loop structure} to solve the original non-convex problem \eqref{P2_eqv_slack}. In the \textbf{inner loop}, the penalty factors $\rho$ and $\lambda$ are held fixed. Utilizing the AO framework, the variables $\left\lbrace x_{m,j}\right\rbrace$ and $\left\lbrace \mathbf \Theta_{\ell,j}\right\rbrace$ are alternately optimized by solving subproblems \eqref{P2_eqv_slack2_sub1_eqv_pena} and \eqref{P2_eqv_slack2_sub2_pen}. Based on the standard complexity result of interior-point methods \cite{2014_K.wang_complexity}, the computational complexity of solving \eqref{P2_eqv_slack2_sub1_eqv_pena} is $\mathcal{O}\left( \left(G + 2MJ\right)^{1.5} (MJ)^2 \ln\left(1/\epsilon\right)\right)$, while solving \eqref{P2_eqv_slack2_sub2_pen} requires a computational complexity of $\mathcal{O}\left( (JN)^2 \left( JN + G\right)^{1.5} \ln\left(1/\epsilon\right) \right)$, where $G \triangleq \sum_{j=1}^J|\mathcal{G}_j|$ and $\epsilon$ denotes the solution accuracy. The AO continues until the joint penalized objective, $\eta - \rho\sum_{m=1}^M\sum_{j=1}^J\left( x_{m,j} - \Pi^{\rm lb, \it r}(x_{m,j})\right) - \lambda \sum_{j\in\mathcal J}\sum_{n\in\mathcal N} \delta_{j,n}$, converges for a given pair of $\{\rho, \lambda\}$. In the \textbf{outer loop}, we evaluate the constraint violations. If the binary or unit-modulus constraints are not satisfied, the penalty factors $\rho$ and $\lambda$ are increased (for instance, by multiplying by a scaling factor $\mu > 1$), and the inner loop is restarted. This procedure continues until the maximum constraint violation falls below a predefined tolerance. 

Finally, we briefly discuss the convergence of the proposed algorithm. With fixed penalty factors, the AO inner loop generates a non-decreasing objective sequence, which is upper-bounded over the feasible set. Hence, the inner loop converges to a stationary point. In the outer loop, the penalty factors $\rho$ and $\lambda$ are progressively increased, driving the penalized constraint violations to vanish as $\rho,\lambda\to\infty$. Consequently, the final solution satisfies the original binary and unit-modulus constraints and provides a feasible starting point for solving the cost minimization problem (P1) in the next section.  %\looseness=-1

\vspace{-3mm}
\section{Proposed Algorithm for Problem (P1)}\label{sec:P1_solution}
In this section, we solve problem (P1). First, analogous to \eqref{P2_opt_w}, the optimal transmit beamformer $\mathbf w_{\mathbf u_j}^{\star}$ is given by
\begin{equation}\label{P1_opt_w}
	\mathbf w_{\mathbf u_j}^{\star} = \frac{\mathbf X_j\left(\sum_{\ell=1}^Lz_\ell\mathbf h_{\ell}^H(\mathbf u_j)\mathbf \Theta_{\ell,j}\overline{\mathbf G}_\ell\right)^H}{\left\| \mathbf X_j\left(\sum_{\ell=1}^Lz_\ell\mathbf h_{\ell}^H(\mathbf u_j)\mathbf \Theta_{\ell,j}\overline{\mathbf G}_\ell\right)^H\right\|}.
\end{equation}
By substituting \eqref{P1_opt_w} into (P1) and adopting the grid-based discretization strategy to handle the semi-infinite SNR constraints (replacing $\mathcal A_j$ with $\mathcal G_j$ as in \eqref{P2_eqv_slack}), we arrive at the following reformulated problem: 
\begin{subequations}\label{P1_eqv_slack}
	\begin{eqnarray}
		&\hspace{-9mm}\underset{\substack{\left\lbrace x_{m,j}\right\rbrace,\left\lbrace z_\ell\right\rbrace,\\\left\lbrace \mathbf \Theta_{\ell,j}\right\rbrace}}{\min} & \hspace{-3mm} c_{\rm MA}\max_{j\in\mathcal J}\left\lbrace \sum_{m=1}^Mx_{m,j}\right\rbrace \! + \! \sum_{\ell=1}^L z_\ell\left( c_\ell +  c_{\rm e} N_\ell\right) \\
		&\hspace{-5mm}\text{s.t.}& \hspace{-4mm} \bar P\left\|\left(\sum_{\ell=1}^Lz_\ell\mathbf h_{\ell}^H(\mathbf u_j)\mathbf \Theta_{\ell,j}\overline{\mathbf G}_\ell\right)\mathbf X_j\right\|^2 \geq \gamma_{\rm th, \it j}, \nonumber\\
		&& \hspace{-4mm}\forall \mathbf u_j \in \mathcal G_j, j\in\mathcal J, \label{P1_eqv_appro_cons:b}\\
		&& \hspace{-4mm} \eqref{P1_cons:d}-\eqref{P1_cons:j}.
	\end{eqnarray}
\end{subequations}
Despite improved tractability, problem \eqref{P1_eqv_slack} remains non-convex due to the coupled variables and the binary/unit-modulus constraints. We tackle these challenges using a double-loop framework similar to that in Section \ref{sec:feasib_check}, initialized with the feasible solution obtained therein. Specifically, the inner loop alternately updates the two variable blocks $\left\lbrace x_{m,j}, z_\ell\right\rbrace$ and $\left\lbrace \mathbf \Theta_{\ell,j}\right\rbrace$ with fixed penalty factors, while the outer loop progressively increases penalties to drive constraint violations to zero. The two subproblems are derived below. 

\vspace{-2mm}
\subsection{Optimizing $\left\lbrace x_{m,j}, z_\ell\right\rbrace$ for Given $\left\lbrace \mathbf \Theta_{\ell,j}\right\rbrace$}
With given $\left\lbrace \mathbf \Theta_{\ell,j}\right\rbrace$, the binary indicators $\left\lbrace x_{m,j}, z_\ell\right\rbrace$ can be jointly optimized by solving the following subproblem of \eqref{P1_eqv_slack}: 
\begin{subequations}\label{P1_eqv_slack_sub1}
	\begin{eqnarray}
		&\hspace{-9mm}\underset{\left\lbrace x_{m,j}\right\rbrace,\left\lbrace z_\ell\right\rbrace}{\min} & \hspace{-3mm} c_{\rm MA}\max_{j\in\mathcal J}\left\lbrace \sum_{m=1}^Mx_{m,j}\right\rbrace + \sum_{\ell=1}^L z_\ell\left( c_\ell +  c_{\rm e} N_\ell\right) \\
		&\hspace{-5mm}\text{s.t.}& \hspace{-4mm}  \eqref{P1_eqv_appro_cons:b}, \eqref{P1_cons:d}-\eqref{P1_cons:i}. 
	\end{eqnarray}
\end{subequations} 
To make the coupling structure in constraint \eqref{P1_eqv_appro_cons:b} more explicit, we expand the term $\left\|\left(\sum_{\ell=1}^Lz_\ell\mathbf h_{\ell}^H(\mathbf u_j)\mathbf \Theta_{\ell,j}\overline{\mathbf G}_\ell\right)\mathbf X_j\right\|^2$ and express it explicitly in terms of $\left\lbrace x_{m,j}, z_\ell\right\rbrace$ as follows: 
\begin{align}\label{SNR_expand}
     & \left\|\left(\sum_{\ell=1}^Lz_\ell\mathbf h_{\ell}^H(\mathbf u_j)\mathbf \Theta_{\ell,j}\overline{\mathbf G}_\ell\right)\mathbf X_j\right\|^2 \nonumber\\
	 & = \sum_{m=1}^Mx_{m,j}\left|\sum_{\ell=1}^Lz_\ell\mathbf h_{\ell}^H(\mathbf u_j)\mathbf \Theta_{\ell,j}\overline{\mathbf G}_\ell(:,m)\right|^2 \nonumber\\
	 & \triangleq \sum_{m=1}^Mx_{m,j}\sum_{\ell=1}^L \sum_{\ell'=1}^L\left(z_\ell b_{\ell,j,m}(\mathbf u_j)\right) \left(z_{\ell'}b_{\ell',j,m}(\mathbf u_j)\right)^* \nonumber\\
	 & \triangleq \sum_{m=1}^Mx_{m,j}\sum_{\ell=1}^L \sum_{\ell'=1}^Lz_\ell z_{\ell'}B_{\ell,\ell',m,j}(\mathbf u_j), 
\end{align}
where $b_{\ell,j,m}(\mathbf u_j) \triangleq \mathbf h_{\ell}^H(\mathbf u_j)\mathbf \Theta_{\ell,j}\overline{\mathbf G}_\ell(:,m) \in \mathbb C$ and $B_{\ell,\ell',m,j}(\mathbf u_j) \triangleq b_{\ell,j,m}(\mathbf u_j)b_{\ell',j,m}^*(\mathbf u_j)$. Accordingly, constraint \eqref{P1_eqv_appro_cons:b} can be equivalently rewritten as
\begin{align}\label{P1_eqv_cons:b_eqv1}
	& \bar P\sum_{m=1}^Mx_{m,j}\sum_{\ell=1}^L \sum_{\ell'=1}^Lz_\ell z_{\ell'}B_{\ell,\ell',m,j}(\mathbf u_j) \geq \gamma_{\rm th, \it j}, \nonumber\\
	& \forall \mathbf u_j \in \mathcal G_j, j\in\mathcal J. 
\end{align}
Note that constraint \eqref{P1_eqv_cons:b_eqv1} involves a non-convex triple-product term $z_\ell z_{\ell'}x_{m,j}$, which couples the binary variables multiplicatively and significantly complicates the optimization. To linearize this constraint, we introduce auxiliary variables:  
\begin{align}
	s_{\ell,\ell',m,j} \triangleq z_\ell z_{\ell'}x_{m,j}, \ \forall \ell,\ell'\in\mathcal L, m\in\mathcal M, j\in\mathcal J,
\end{align}
and impose the following additional constraints:
\begin{subequations}\label{s_cons}
    \begin{align}
    	& s_{\ell,\ell',m,j} \in \{0,1\}, \ \forall \ell,\ell'\in\mathcal L, m\in\mathcal M, j\in\mathcal J, \label{s_cons:a}\\
    	& s_{\ell,\ell',m,j} \leq z_\ell, \  s_{\ell,\ell',m,j} \leq z_{\ell'}, \ s_{\ell,\ell',m,j} \leq x_{m,j}, \nonumber\\
    	& \forall \ell,\ell'\in\mathcal L, m\in\mathcal M, j\in\mathcal J, \label{s_cons:b}\\
    	& s_{\ell,\ell',m,j} \geq z_\ell + z_{\ell'} + x_{m,j} - 2, \ \forall \ell,\ell'\in\mathcal L, m\in\mathcal M, j\in\mathcal J. \label{s_cons:c}
    \end{align}
\end{subequations}
It can be readily verified that, under the binary constraints \eqref{P1_cons:e} and \eqref{P1_cons:h}, as well as the linearization conditions in \eqref{s_cons}, constraint \eqref{P1_eqv_cons:b_eqv1} is equivalent to
\begin{align}\label{P1_eqv_cons:b_eqv2}
	&\bar P\sum_{m=1}^M\sum_{\ell=1}^L\sum_{\ell'=1}^Ls_{\ell,\ell',m}B_{\ell,\ell',m,j}(\mathbf u_j) \geq \gamma_{\rm th, \it j}, \nonumber\\
	&\forall \mathbf u_j \in \mathcal G_j, j\in\mathcal J.
\end{align}
By replacing constraint \eqref{P1_eqv_cons:b_eqv1} with \eqref{P1_eqv_cons:b_eqv2} and taking the constraints in \eqref{s_cons} into account, we arrive at the following equivalent form of problem \eqref{P1_eqv_slack_sub1}: 
\begin{subequations}\label{P1_eqv_slcak_sub1_eqv}
	\begin{eqnarray}
		&\hspace{-8mm}\underset{\substack{\{x_{m,j}\},\{z_\ell\},\\ \{s_{\ell,\ell',m,j}\}}}{\min}& \hspace{-3mm} c_{\rm MA}\max_{j\in\mathcal J}\left\lbrace \sum_{m=1}^Mx_{m,j}\right\rbrace \! + \! \sum_{\ell=1}^L z_\ell\left( c_\ell +  c_{\rm e} N_\ell\right) \\
		&\hspace{-7mm}\text{s.t.}& \hspace{-5mm} \eqref{P1_eqv_cons:b_eqv2}, \eqref{s_cons},  \eqref{P1_cons:e}-\eqref{P1_cons:i}. 
	\end{eqnarray}
\end{subequations} 
The remaining obstacle to solving this problem is the binary constraints on $\left\lbrace x_{m,j}, z_\ell, s_{\ell,\ell',m,j}\right\rbrace$. To tackle this issue, we employ the same penalty method as in Section \ref{sec:feasib_check_A}. Note that the auxiliary variables $\left\lbrace s_{\ell,\ell',m,j} \right\rbrace$ are coupled with the binary variables $\left\lbrace x_{m,j}, z_\ell\right\rbrace$ through the linear inequalities in \eqref{s_cons}, which constitute a tight convex-hull linearization of the trilinear product $s_{\ell,\ell',m,j} = z_\ell z_{\ell'}x_{m,j}$ over the unit hypercube. Specifically, when $\left\lbrace x_{m,j}, z_\ell\right\rbrace \in \{0,1\}$, these linear inequalities inherently force $\left\lbrace s_{\ell,\ell',m,j} \right\rbrace$ to take binary values that match the corresponding products.  Hence, it suffices to penalize only the binary-constraint violations of $\left\lbrace x_{m,j}, z_\ell\right\rbrace$. 

Accordingly, we relax $\left\lbrace x_{m,j}, z_\ell, s_{\ell,\ell',m,j}\right\rbrace$ from $\{0,1\}$ to $[0,1]$. Since the SCA-based penalty construction has been detailed in Section~\ref{sec:feasib_check_A}, we omit the derivations and directly apply the same linearization to the quadratic terms $x_{m,j}^2$ and $z_\ell^2$. Let $\Lambda^{\rm lb, \it r}(z_\ell) \triangleq (z_\ell^r)^2 + 2z_\ell^r(z_\ell - z_\ell^r)$ denote the first-order Taylor expansion of $z_\ell^2$ at a given local point $z_\ell^r$, and let $\Pi^{\rm lb, \it r}(x_{m,j})$ follow the definition in \eqref{ineq:x_lower_bound}. By adding the corresponding penalty terms for $\left\lbrace x_{m,j},z_\ell\right\rbrace$ to the objective, we obtain the following tractable convex subproblem: 
\begin{subequations}\label{P1_eqv_slcak_sub1_final}
	\begin{eqnarray}
		&\hspace{-8mm}\underset{\substack{\{x_{m,j}\},\{z_\ell\},\\ \{s_{\ell,\ell',m,j}\}}}{\min}& \hspace{-1.5mm} c_{\rm MA}\max_{j\in\mathcal J}\left\lbrace \sum_{m=1}^Mx_{m,j}\right\rbrace + \sum_{\ell=1}^L z_\ell\left( c_\ell +  c_{\rm e} N_\ell\right) \nonumber\\
		&& + \zeta \sum_{j=1}^J\sum_{m=1}^M \left( x_{m,j} - \Pi^{\rm lb, \it r}(x_{m,j}) \right) \nonumber\\
		&& + \zeta \sum_{\ell=1}^L \left( z_\ell - \Lambda^{\rm lb, \it r}(z_\ell) \right) \\
		&\hspace{-7mm}\text{s.t.}& \hspace{-5mm} 0 \le x_{m,j} \le 1, \ 0 \le z_\ell \le 1, \ 0 \le s_{\ell,\ell',m,j} \le 1, \nonumber\\
		&& \hspace{-5mm} \forall m\in\mathcal M, j\in\mathcal J, \ell,\ell'\in\mathcal L, \\
		&& \hspace{-5mm} \eqref{P1_eqv_cons:b_eqv2}, \eqref{s_cons:b},\eqref{s_cons:c}, \eqref{P1_cons:f}, \eqref{P1_cons:g}, \eqref{P1_cons:i}.
	\end{eqnarray}
\end{subequations} 
Problem \eqref{P1_eqv_slcak_sub1_final} is a convex linear program that can be efficiently solved using off-the-shelf solvers (e.g., CVX). Driven by the increasing penalty factor $\zeta$, the primary variables $\left\lbrace x_{m,j}, z_\ell\right\rbrace$ converge to binary values, which in turn forces the auxiliary variables $\left\lbrace s_{\ell,\ell',m,j}\right\rbrace$ to satisfy the binary requirement due to the tightness of the inequalities in \eqref{s_cons:b} and \eqref{s_cons:c}.

\subsection{Optimizing $\left\lbrace \mathbf \Theta_{\ell,j}\right\rbrace$ for Given $\left\lbrace x_{m,j}, z_\ell\right\rbrace$}
With the deployment configuration $\left\lbrace x_{m,j}, z_\ell\right\rbrace$ fixed, problem \eqref{P1_eqv_slack} reduces to a feasibility problem regarding the IRS phase-shift variables $\left\lbrace \mathbf \Theta_{\ell,j}\right\rbrace$. To improve the convergence behavior, we introduce slack variables $\{\beta_j\}$ to quantify the per-area SNR margins and reformulate the feasibility task as a margin-maximization problem. These slack variables explicitly represent the SNR margin for each target area $j$. Maximizing these margins enlarges the feasible set for the subsequent update of $\left\lbrace x_{m,j}, z_\ell\right\rbrace$, which helps promote sparser deployments (i.e., fewer deployed MAs/IRSs) in the next iteration. Accordingly, the subproblem for optimizing $\left\lbrace \mathbf \Theta_{\ell,j}\right\rbrace$ is reformulated as 
\begin{subequations}\label{P1_eqv_slack_sub2}
	\begin{eqnarray}
		&\hspace{-2mm}\underset{\left\lbrace \mathbf \Theta_{\ell,j}\right\rbrace, \left\lbrace \beta_j \geq 0\right\rbrace}{\max} & \sum_{j=1}^J \beta_j \\
		&\hspace{-2mm}\text{s.t.}& \hspace{-9mm} \bar P\left\|\left(\sum_{\ell=1}^Lz_\ell\mathbf h_{\ell}^H(\mathbf u_j)\mathbf \Theta_{\ell,j}\overline{\mathbf G}_\ell\right)\mathbf X_j\right\|^2 \geq \gamma_{\rm th, \it j} + \beta_j, \nonumber\\
		&& \hspace{-9mm}\forall \mathbf u_j \in \mathcal G_j, j\in\mathcal J, \label{P1_eqv_slack_sub2_cons:b}\\
		&& \hspace{-9mm} \eqref{P1_cons:j}.
	\end{eqnarray}
\end{subequations}
Problem \eqref{P1_eqv_slack_sub2} optimizes the phase shifts only for the deployed IRSs with $z_\ell=1$. Define the deployed-IRS index set $\mathcal L_{\rm s} \triangleq \left\lbrace\ell: z_\ell = 1\right\rbrace \subseteq \mathcal L$, and let $\mathcal N_{\rm s}$ collect the indices of their reflecting elements, with $N_{\rm s} \triangleq \left|\mathcal N_{\rm s}\right| $. Since the quadratic term $\left\|\left(\sum_{\ell=1}^Lz_\ell\mathbf h_{\ell}^H(\mathbf u_j)\mathbf \Theta_{\ell,j}\overline{\mathbf G}_\ell\right)\mathbf X_j\right\|^2$ has the same structure as in \eqref{quad_term_eqv}, we reuse the compact notations $\mathbf v_j$ and $\mathbf R(\mathbf u_j)$, with dimensions defined over $\mathcal L_{\rm s}$ and $\mathcal N_{\rm s}$. Consequently, problem \eqref{P1_eqv_slack_sub2} can be recast as  
\begin{subequations}\label{P1_eqv_slack_sub2_eqv}
	\begin{eqnarray}
		&\hspace{-9mm}\underset{\left\lbrace \mathbf v_j\right\rbrace, \left\lbrace \beta_j \geq 0 \right\rbrace}{\max} & \sum_{j=1}^J \beta_j \\
		&\hspace{-9.5mm}\text{s.t.}& \hspace{-8mm} \bar P\mathbf v_j^H\mathbf R(\mathbf u_j)\mathbf v_j \geq \gamma_{\rm th, \it j} + \beta_j, \forall \mathbf u_j \in \mathcal G_j, j\in\mathcal J, \label{P1_eqv_slack_sub2_cons:b_eqv}\\
		&& \hspace{-8mm} \left|\left[\mathbf v_j\right]_n\right| = 1, \ \forall  j\in\mathcal J, n\in\mathcal N_{\rm s}. \label{P1_eqv_slack_sub2_eqv_cons:c}
	\end{eqnarray}
\end{subequations}
Given that problem \eqref{P1_eqv_slack_sub2_eqv} shares the same mathematical structure as \eqref{P2_eqv_slack2_sub2} in Section \ref{sec:feasib_check_B}, we apply the identical SCA and penalty strategies. For brevity, we omit repetitive derivations and directly formulate the convex surrogate subproblem as: 
\begin{subequations}\label{P1_eqv_slack_sub2_eqv_pen}
	\begin{eqnarray}
		&\hspace{-8mm}\underset{\substack{\{\mathbf v_j\}, \{\beta_j\}, \\ \{\delta_{j,n}\ge 0\}}}{\max} & \sum_{j=1}^J \beta_j - \xi \sum_{j=1}^J\sum_{n=1}^{N_{\rm s}} \delta_{j,n} \\
		&\hspace{-1cm}\text{s.t.}& \hspace{-8mm} \bar P \Xi^{\rm lb,\it r}(\mathbf v_j, \mathbf u_j) \geq \gamma_{\rm th, \it j} + \beta_j, \ \forall \mathbf u_j \in \mathcal G_j, j\in\mathcal J, \\
		&& \hspace{-8mm} \left|\left[\mathbf v_j\right]_n\right|^2 \le 1,\ \forall j\in\mathcal J, n\in\mathcal N_{\rm s}, \\
		&& \hspace{-8mm} 1-\delta_{j,n}\le \chi^{\rm lb,\it r}\!\left(\left[\mathbf v_j\right]_n\right), \ \forall j\in\mathcal J, n\in\mathcal N_s,
	\end{eqnarray}
\end{subequations}
where $\xi$ is a penalty factor, and the linearized terms $\Xi^{\rm lb,\it r}(\cdot)$ and $\chi^{\rm lb,\it r}(\cdot)$ follow the definitions in Section \ref{sec:feasib_check_B}. Problem \eqref{P1_eqv_slack_sub2_eqv_pen} is a convex QCQP that can be efficiently solved by existing convex optimization solvers such as CVX.

\subsection{Overall Algorithm} 
The proposed algorithm for solving problem \eqref{P1_eqv_slack} adopts a similar double-loop structure to that in Section \ref{sec:feasib_check}. Specifically, the inner loop alternates between optimizing the integer block $\left\lbrace x_{m,j}, z_\ell\right\rbrace$ and the phase-shift block $\left\lbrace \mathbf \Theta_{\ell,j}\right\rbrace$, while the penalty factors are updated in the outer loop. Given that the algorithmic framework and convergence properties are analogous to those in the previous section, we omit redundant details and focus solely on the computational complexity, which varies due to the expanded set of variables. 

The computational burden is dominated by the resolution of two convex subproblems within each inner iteration. Specifically, solving subproblem \eqref{P1_eqv_slcak_sub1_final} entails a complexity of $\mathcal{O}\left(C_{\rm dep} \ln\left(1/\epsilon\right)\right)$, where $C_{\rm dep} \triangleq \left(L^2MJ\right)^2\left(L^2MJ+G\right)^{1.5}$, while solving subproblem \eqref{P1_eqv_slack_sub2_eqv_pen} requires $\mathcal{O}\left(\mathcal C_{\rm ph}\ln\left(1/\epsilon\right)\right)$, with $\mathcal C_{\rm ph} \triangleq\sqrt{J N_{\rm s} + G} J N_{\rm s} \left(J^2 N_{\rm s}^2 + J N_{\rm s} G + G^2\right)$ \cite{2014_K.wang_complexity}. Therefore, the per-iteration complexity of the inner loop is $\mathcal O\left(\left(  C_{\rm dep} + \mathcal C_{\rm ph}\right)\ln\left(1/\epsilon\right)\right)$. 

\begin{rem}
	\rm The proposed penalty-based double-loop algorithm readily extends to the \textbf{budget-constrained worst-case SNR maximization} counterpart. Specifically, by introducing an auxiliary variable to represent the worst-case SNR and maximizing it subject to a linear cost constraint, the resulting formulation preserves the mathematical structure and non-convex coupling of problem (P1). Consequently, the AO steps and penalty-based SCA iterations remain effective in handling the binary deployment decisions and non-convex constraints with only minor modifications. %\looseness=-1
\end{rem}

\section{IRS Element Pruning for Further Cost Reduction}\label{sec:element_sizing}
In the preceding formulation, candidate site $\ell$ corresponds to an IRS with a predetermined size $N_\ell$. Thus, selecting site $\ell$ (i.e., $z_\ell=1$) incurs a fixed element-related cost $c_{\rm e} N_\ell$. This predetermined sizing can be conservative, often yielding an achieved SNR above the target. To exploit this potential margin, we introduce a secondary refinement stage: given the obtained $\left\lbrace x_{m,j}, z_\ell, \mathbf\Theta_{\ell,j}\right\rbrace$, we optimize the number of installed elements at each selected IRS to reduce the element-related cost while preserving the SNR guarantees. 

%\vspace{-3mm}
\subsection{Problem Formulation} 
We introduce a binary installation indicator $y_{\ell,n}\in\{0,1\}$ for the $n$-th reflecting element at candidate IRS site $\ell$, where $y_{\ell,n}=1$ indicates that this element is installed and $y_{\ell,n}=0$ otherwise. Define $\mathbf Y_\ell = \mathrm{diag}\left(y_{\ell,1},\ldots,y_{\ell,N_\ell} \right)$, $\forall \ell\in\mathcal L$. Then, the received SNR at location $\mathbf u_j$ in \eqref{eq:SNR} is re-expressed as
\begin{align}\label{eq:SNR_ele}
	\gamma_{\mathbf u_j}
	= \bar P\left|\left(\sum_{\ell=1}^L z_\ell \mathbf h_{\ell}^H(\mathbf u_j)\mathbf Y_\ell \mathbf \Theta_{\ell,j}\overline{\mathbf G}_\ell\right)\mathbf X_j\mathbf w_{\mathbf u_j}\right|^2.
\end{align}
For any given $\{\mathbf Y_\ell\}$, the MRT beamformer remains optimal. As a result, \eqref{eq:SNR_ele} reduces to
\begin{align}\label{eq:SNR_ele_MRT}
	\gamma_{\mathbf u_j}
	= \bar P \left\|\left(\sum_{\ell=1}^L z_\ell \mathbf h_{\ell}^H(\mathbf u_j)\mathbf Y_\ell \mathbf \Theta_{\ell,j}\overline{\mathbf G}_\ell\right)\mathbf X_j\right\|^2.
\end{align}
Moreover, the total infrastructure deployment cost is re-expressed as
\begin{align}
	C_{\rm tot} = \sum_{\ell=1}^L z_\ell c_{\rm e}\sum_{n=1}^{N_\ell}y_{\ell,n} + C_{\rm fixed}, 
\end{align}
where $C_{\rm fixed} \triangleq c_{\rm MA}\max_{j\in\mathcal J}\left\lbrace \sum_{m=1}^Mx_{m,j}\right\rbrace + \sum_{\ell=1}^L z_\ell c_\ell$. 

We aim to minimize the element-related deployment cost by optimizing the installation indicators $\{y_{\ell,n}\}$, while maintaining the SNR requirements. This yields the following problem: 
\begin{subequations}\label{P3}
	\begin{eqnarray}
		&\hspace{-8mm}\text{(P3)}:& \hspace{-1mm}\underset{\left\lbrace y_{\ell,n}\right\rbrace }{\min} \hspace{2mm} \sum_{\ell=1}^L z_\ell c_{\rm e} \sum_{n=1}^{N_\ell}y_{\ell,n} + C_{\rm fixed} \\
		&\hspace{-8mm}\text{s.t.}& \hspace{-3mm}  \bar P \left\|\left(\sum_{\ell=1}^L z_\ell \mathbf h_{\ell}^H(\mathbf u_j)\mathbf Y_\ell \mathbf \Theta_{\ell,j}\overline{\mathbf G}_\ell\right)\mathbf X_j\right\|^2 \geq \gamma_{\rm th, \it j}, \nonumber\\
		&& \hspace{-3mm} \forall u_j \in \mathcal G_j, j\in\mathcal J, \label{P3_cons:b}\\
		&& \hspace{-3mm}  \mathbf Y_\ell = \mathrm{diag}\left(y_{\ell,1},\ldots,y_{\ell,N_\ell} \right),\ \forall \ell\in\mathcal L, \label{P3_cons:c}\\
		&& \hspace{-3mm} y_{\ell,n} \in \{0,1\}, \ \forall \ell\in\mathcal L, n\in\mathcal N_\ell, \label{P3_cons:d}\\
		&& \hspace{-3mm} y_{\ell,n} \leq z_\ell,\ \forall \ell\in\mathcal L, n\in\mathcal N_\ell, \label{P3_cons:e}
	\end{eqnarray}
\end{subequations} 
where constraint \eqref{P3_cons:e} implies that $y_{\ell,n}=0$ whenever $z_\ell=0$. Problem (P3) is a pure binary optimization problem. The binary constraints render it combinatorial, and the quadratic SNR constraints further make the feasible set non-convex, which together make (P3) challenging to solve globally. 

\subsection{Proposed Algorithm for Problem (P3)}
To efficiently solve (P3), we first reformulate the quadratic constraints into a more tractable form. Let  $\mathbf f_{\ell,j,n}^T$ denote the $n$-th row of the effective channel matrix $\mathbf F_{\ell,j} \triangleq \mathbf \Theta_{\ell,j}\overline{\mathbf G}_\ell \mathbf X_j \in \mathbb C^{N_\ell \times M}$. By defining an equivalent aggregated channel vector $\mathbf c_{\ell,j,n}^T(\mathbf u_j) \triangleq \left[\mathbf h_{\ell}^H(\mathbf u_j)\right]_n\!\mathbf f_{\ell,j,n}^T$, the beamforming gain term in \eqref{P3_cons:b} can be expanded as
\begin{align}
	\gamma_{\mathbf u_j} & = \bar P\left\|\sum_{\ell=1}^L \sum_{n=1}^{N_\ell} z_\ell y_{\ell,n}\!\left[\mathbf h_{\ell}^H(\mathbf u_j)\right]_n\!\mathbf f_{\ell,j,n}^T\right\|^2\nonumber\\
	& = \bar P \left\| \sum_{\ell=1}^L \sum_{n=1}^{N_\ell} z_\ell y_{\ell,n}\,\mathbf c_{\ell,j,n}^T(\mathbf u_j) \right\|^2 , \nonumber\\
	& = \bar P \sum_{\ell,n}\sum_{\ell',n'} z_\ell z_{\ell'}\, y_{\ell,n} y_{\ell',n'} \, \mathbf c_{\ell,j,n}^T(\mathbf u_j) \mathbf c_{\ell',j,n'}^*(\mathbf u_j).
\end{align}
For compact representation, we stack all binary variables $\{y_{\ell,n}\}$ into a single vector $\mathbf y \triangleq [\mathbf y_1^T, \mathbf y_2^T, \ldots, \mathbf y_L^T]^T \in \{0,1\}^{N\times 1}$. We then construct a Hermitian matrix $\mathbf Q(\mathbf u_j) \in \mathbb C^{N\times N}$, where the entry corresponding to the indices of $\left(\ell_u, n_u\right)$ and $\left(\ell_v, n_v\right)$ is given by
\begin{equation}
	\left[ \mathbf Q(\mathbf u_j)\right]_{u,v} = z_{\ell_u} z_{\ell_v}\mathbf{c}_{\ell_u, j, n_u}^{T}(\mathbf u_j)\mathbf{c}_{\ell_v, j, n_v}^{*}(\mathbf u_j). 
\end{equation}
Accordingly, $\gamma_{\mathbf u_j}$ is recast as the quadratic form $\gamma_{\mathbf u_j} = \bar P \mathbf y^H \mathbf Q(\mathbf u_j) \mathbf y$. Problem (P3) is equivalently transformed into 
\begin{subequations}\label{P3_eqv}
	\begin{eqnarray}
		&\underset{\left\lbrace y_{\ell,n}\right\rbrace }{\min}& \sum_{\ell=1}^L z_\ell c_{\rm e} \sum_{n=1}^{N_\ell}y_{\ell,n} + C_{\rm fixed} \\
		&\text{s.t.}& \hspace{-2mm}  \bar P \mathbf y^H \mathbf Q(\mathbf u_j) \mathbf y \geq \gamma_{\rm th, \it j}, \  \forall u_j \in \mathcal G_j, j\in\mathcal J, \label{P3_eqv_cons:b}\\
		&& \hspace{-2mm} \eqref{P3_cons:d}, \eqref{P3_cons:e}.  
	\end{eqnarray}
\end{subequations} 
To address the non-convex constraint \eqref{P3_eqv_cons:b}, we adopt the SCA method. Specifically, the quadratic term $\mathbf y^H \mathbf Q(\mathbf u_j) \mathbf y$ is replaced by its first-order Taylor expansion-based lower bound at a given local point $\mathbf y^r$ in the $r$-th iteration. As a result, problem \eqref{P3_eqv} is approximated as
\begin{subequations}\label{P3_eqv_sca}
	\begin{eqnarray}
		&\hspace{-2.5mm}\underset{\left\lbrace y_{\ell,n}\right\rbrace }{\min}& \sum_{\ell=1}^L z_\ell c_{\rm e} \sum_{n=1}^{N_\ell}y_{\ell,n} + C_{\rm fixed} \\
		&\hspace{-2.5mm}\text{s.t.}& \hspace{-4mm}  \bar P \left( 2\Re\left\lbrace \left(\mathbf y^r\right)^H\mathbf Q(\mathbf u_j)\mathbf y\right\rbrace - \left(\mathbf y^r\right)^H\mathbf Q(\mathbf u_j)\mathbf y^r\right) \geq \gamma_{\rm th, \it j}, \nonumber\\
		&& \hspace{-4mm} \forall u_j \in \mathcal G_j, j\in\mathcal J, \label{P3_eqv_sca_cons:a}\\
		&& \hspace{-4mm} \eqref{P3_cons:d}, \eqref{P3_cons:e}.  
	\end{eqnarray}
\end{subequations} 
Problem \eqref{P3_eqv_sca} is a standard 0-1 integer linear program. To avoid combinatorial search, we relax the binary variables $\left\lbrace y_{\ell,n}\right\rbrace$ from $\{0,1\}$ to $[0,1]$ and add an integrality-violation penalty to the objective, thereby transforming problem \eqref{P3_eqv_sca} into the following convex linear program:
\begin{subequations}\label{P3_eqv_sca_pen}
	\begin{eqnarray}
		&\underset{\left\lbrace y_{\ell,n}\right\rbrace }{\min}& \sum_{\ell=1}^L z_\ell c_{\rm e} \sum_{n=1}^{N_\ell}y_{\ell,n} + C_{\rm fixed} \nonumber\\
		&& + \nu\sum_{\ell=1}^L \sum_{n=1}^{N_\ell} \left( y_{\ell,n} - \Upsilon^{\rm lb, \it r}(y_{\ell,n}) \right) \\
		&\text{s.t.}& \hspace{-2mm}  0 \le y_{\ell,n} \le 1, \ \forall \ell\in\mathcal L, n\in\mathcal N_\ell,  \\
		&& \hspace{-2mm} \eqref{P3_eqv_sca_cons:a}, \eqref{P3_cons:e}, 
	\end{eqnarray}
\end{subequations}
where $\nu\geq 0$ is a penalty factor, and $\Upsilon^{\mathrm{lb},r}\!\left(y_{\ell,n}\right) \triangleq \left(y_{\ell,n}^{r}\right)^{2} + 2y_{\ell,n}^{r}\left(y_{\ell,n}-y_{\ell,n}^{r}\right)$ represents the first-order Taylor expansion of $y_{\ell,n}^2$ at a given local point $y_{\ell,n}^r$. Problem \eqref{P3_eqv_sca_pen} can be efficiently solved using convex optimization tools such as CVX, with a complexity of $\mathcal{O}\left( (G + 2N)^{1.5} N^2 \ln(1/\epsilon) \right)$ \cite{2014_K.wang_complexity}. The final binary solution is obtained by iteratively solving problem \eqref{P3_eqv_sca_pen} with a progressively increasing penalty factor $\nu$. \looseness=-1

\section{Simulation Results}\label{sec:simulation}
This section evaluates the proposed designs through numerical simulations. The carrier frequency is 3 GHz ($\lambda = 0.1$ m) \cite{2024_Zhenyu_uplink}, giving a reference channel gain at 1 m of $C_0 = (\lambda/4\pi)^2$. We consider $L = 5$ candidate IRS sites with coordinates (in m): $[5,0,12]^T$, $[0,12,5]^T$, $[0,-12,5]^T$, $[10,25,5]^T$, and $[10,-25,5]^T$. The first site hosts a horizontally oriented IRS facing downwards (e.g., on an uncrewed aerial vehicle or a ceiling), while the others are vertically oriented (e.g., on building facades or walls). Each IRS is a uniform planar array with half-wavelength spacing. Target areas are $5~\mathrm{m} \times 5~\mathrm{m}$ disjoint squares, randomly placed within $x \in [50, 70]$ m, $y \in [-40, 40]$ m, $z = 0$ m, and sampled at 1-m intervals. Cost parameters are normalized to the per-element cost $c_{\rm e} = 1$. The fixed deployment costs for the five sites are given by $c_{\ell} = \{30,20,20,10,10\}$ for $\ell = 1,\ldots,5$. Default simulation parameters are: $P = 20$ dBm, $\sigma^2 = -90$ dBm, $J = 2$, $c_{\rm MA} = 30$, $A = 3\lambda$, $D = \lambda/2$  \cite{2023_Wenyan_MIMO}, $d = \lambda/2$, and $N_\ell = 50$ potential elements per IRS (arranged $5\times 10$). A uniform SNR target $\gamma = 10$ dB is imposed for all $\mathbf u_j \in \mathcal G_j, j\in\mathcal J$. %\looseness=-1

\begin{figure}[!t]
	\vspace{-3mm}
	\centering
	\includegraphics[scale=0.65]{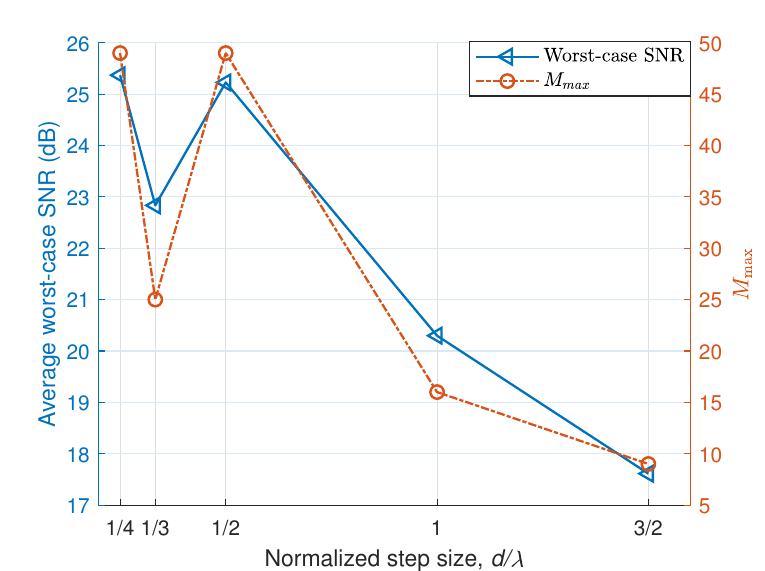}
	\caption{Average worst-case SNR versus normalized step size.} \label{Fig:feasibility_vs_d}
	\vspace{-3mm}
\end{figure}

\vspace{-1mm}
\subsection{Feasibility Check via Worst-Case SNR Maximization}
In this subsection, we evaluate the feasibility limit under the full-configuration scenario. With identical target SNRs, maximizing $\eta$ in the feasibility check problem (P2) is mathematically equivalent to maximizing the worst-case SNR across all the target areas. We therefore use the maximum achievable worst-case SNR as our primary metric. This value provides a quantitative feasibility boundary: any target SNR below it is feasible, offering more information than a binary feasible–infeasible check. %\looseness=-1

Fig.~\ref{Fig:feasibility_vs_d} shows the average worst-case SNR (left axis) and the maximum allowable number of MAs, $M_{\max}$ (right axis), versus the normalized grid step size $d/\lambda$. The worst-case SNR closely tracks $M_{\max}$, which is determined by the interplay between the grid step size $d$ and the minimum inter-MA spacing requirement $D$. In particular, $d=\lambda/4$ and $d=\lambda/2$ both enable near-optimal packing ($M_{\max} = 49$), yielding the highest SNR of about $25$~dB. Although $d=\lambda/4$ provides slightly more spatial flexibility, its gain over $d=\lambda/2$ is marginal because $M_{\max}$ is saturated in both cases. By contrast, a pronounced dip occurs at $d=\lambda/3$, where grid-spacing misalignment limits packing ($M_{\max} = 25$) and reduces the resulting SNR to approximately $23$~dB, indicating that a finer grid does not necessarily improve performance. For coarse grids ($d>\lambda/2$), $M_{\max}$ decreases monotonically, leading to a substantial SNR loss. Overall, $d=\lambda/2$ offers a good trade-off between performance and computational complexity. 

\begin{figure*}[!ht]
	\vspace{-3mm}
	\centering
	%\hspace{-4mm}
	\subfigure[]{\label{Fig:cost_vs_cMA}
	\includegraphics[scale=0.65]{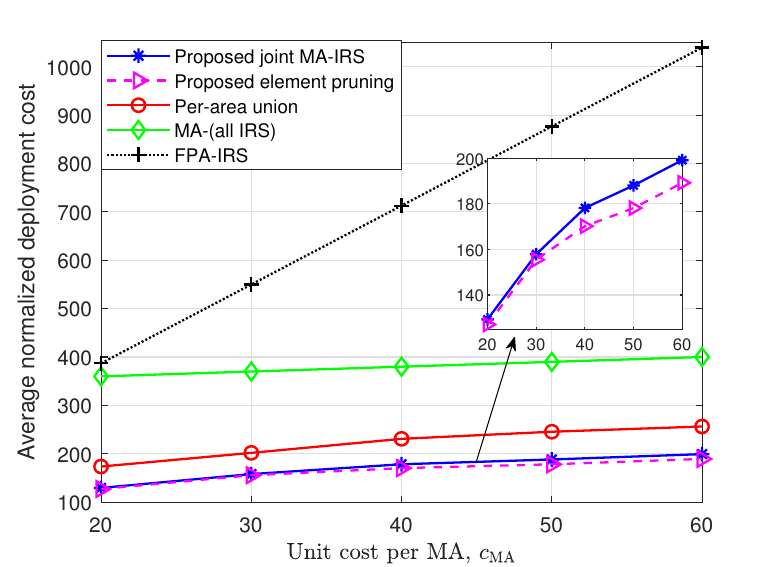}}
	%\hspace{-4mm}
	\subfigure[]{\label{Fig:x_z_vs_cMA}
		\includegraphics[scale=0.65]{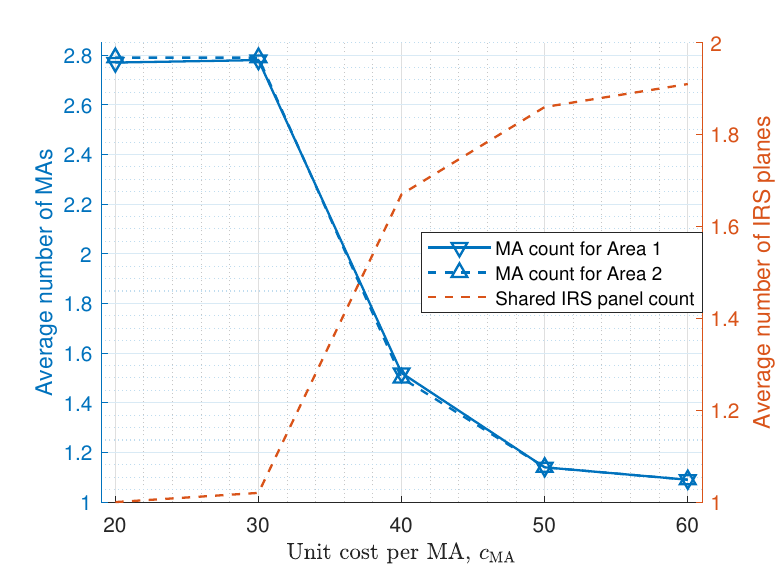}}
	\caption{Impact of the MA unit cost $c_{\rm MA}$. (a) Average normalized deployment cost. (b) Average MA count per area and shared IRS panel count under the proposed joint MA-IRS design.}
	\label{Fig:cost_x_z_vs_cMA}
\end{figure*}

\vspace{-2mm}
\subsection{Deployment Cost Minimization Under SNR Constraints}

In this subsection, we assess the cost-effectiveness of the proposed designs by comparing the following schemes. For all schemes, the transmit beamforming is chosen as MRT based on the end-to-end effective channels.
\begin{itemize}
	\item \textbf{Proposed joint MA-IRS:} The approach in Section~\ref{sec:P1_solution}, which jointly optimizes the MA placement, IRS site selection, and IRS phase shifts, assuming full-size IRS panels (i.e., $N_\ell$ elements) at the selected sites. 
	
	\item \textbf{Proposed element pruning:} The method in Section~\ref{sec:element_sizing} that further optimizes the number of installed IRS elements based on the joint MA-IRS solution. 
	
	\item \textbf{Per-area union:} A baseline that performs independent optimization for each target area, with the deployment cost computed from the union of selected IRS sites and the maximum MA number over all areas.  
	
	\item \textbf{MA-(all IRS):} A benchmark with all candidate IRSs deployed, i.e., $z_\ell = 1,\ \forall \ell \in \mathcal L$, where only the MA placement and IRS phase shifts are optimized.
	
	\item \textbf{FPA-IRS:} An FPA benchmark that deploys antennas at all feasible grid points (subject to the minimum spacing constraint) and optimizes the IRS site selection and phase shifts. The unit cost ratio is defined as $\kappa \triangleq c_{\rm FPA}/c_{\rm MA}$ and is set to $\kappa = 1/3$ unless otherwise specified.
\end{itemize}

\subsubsection{Impact of Unit Cost of MAs}
Fig.~\ref{Fig:cost_x_z_vs_cMA} examines the impact of the MA unit cost $c_{\rm MA}$ on the deployment performance. In Fig.~\ref{Fig:cost_vs_cMA}, the average normalized deployment cost increases with $c_{\rm MA}$ for all schemes, with the FPA-IRS scheme rising the fastest because its fully populated FPA array incurs a large fixed antenna cost even when $\kappa=1/3$. The MA-(all IRS) benchmark remains costly yet relatively insensitive to $c_{\rm MA}$ because deploying all $5$ IRSs dominates the cost. The per-area union baseline performs better but still incurs higher costs than the proposed methods due to the lack of inter-area resource sharing. By contrast, the proposed joint MA-IRS scheme achieves substantial cost savings by jointly optimizing MA placement and selecting only the necessary IRSs. As highlighted in the zoomed-in view, the proposed element pruning scheme further reduces the cost by pruning redundant elements from the selected IRSs, validating the benefit of the second-stage refinement.

To provide deeper insights into the cost-saving mechanism of the proposed joint MA-IRS scheme, Fig.~\ref{Fig:x_z_vs_cMA} plots the average number of deployed MAs and shared IRS panels versus $c_{\rm MA}$. A clear resource trade-off is observed. In the low-cost regime where $c_{\rm MA} \le 30$, the system favors deploying more MAs, approximately $2.8$ per area, while activating only a minimal number of IRSs, around $1$. This occurs because MAs provide high beamforming gain at a low cost, which reduces the reliance on IRSs. However, as $c_{\rm MA}$ increases beyond $30$, the number of MAs drops sharply to nearly $1$. Meanwhile, the number of deployed IRSs increases to compensate for the reduced active gain. This demonstrates the capability of the algorithm to adaptively substitute expensive active resources with more cost-effective passive resources to minimize the total deployment cost.

\begin{figure}[!t]
	\vspace{-3.5mm}
	\centering
	\includegraphics[scale=0.65]{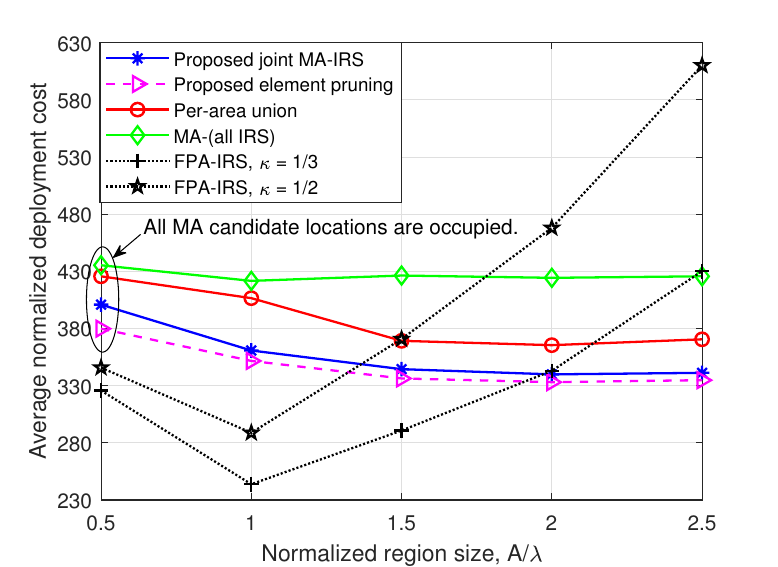}
	\caption{Average normalized deployment cost versus normalized region size.} \label{Fig:cost_vs_reg}
	\vspace{-3mm}
\end{figure}

\begin{figure}[!t]
	\vspace{-3.5mm}
	\centering
	\includegraphics[scale=0.65]{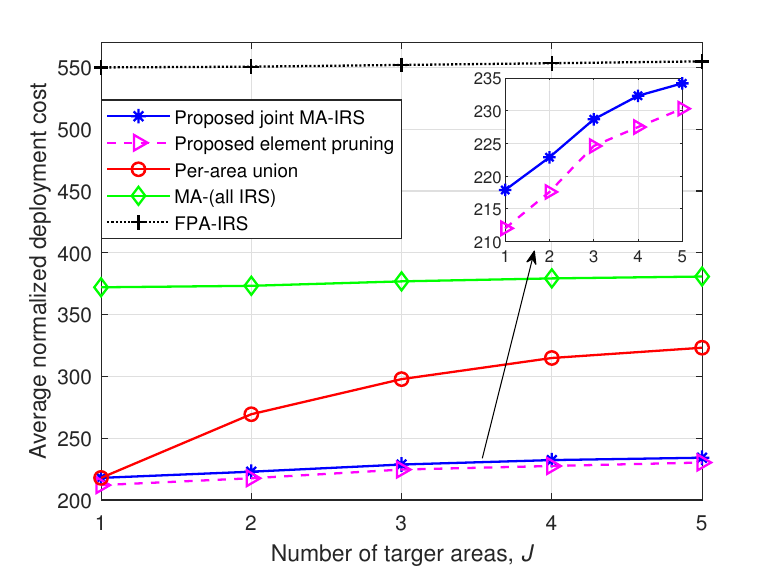}
	\caption{Average normalized deployment cost versus number of target areas.} \label{Fig:cost_vs_J}
	\vspace{-3mm}
\end{figure}

\subsubsection{Impact of Normalized Region Size}
Fig.~\ref{Fig:cost_vs_reg} illustrates the deployment cost versus the normalized region size $A/\lambda$, where the target SNR is set to $\gamma = 15$~dB. Two notable phenomena regarding the cost behaviors are observed. First, the cost of the FPA-IRS scheme follows a non-monotonic U-shaped trend, which decreases initially before escalating rapidly. The initial decrease occurs because the enhanced array gain from a slightly larger aperture allows the system to deactivate expensive IRSs. The resulting savings outweigh the cost of adding a few cheap FPA elements. However, as $A$ grows further, the cost rises sharply since the number of fixed antennas grows quadratically. Second, the proposed joint MA-IRS scheme is not invariably superior to the FPA benchmarks, especially in compact regions. At $A=0.5\lambda$, the MA system is saturated and physically degenerates into a fully populated array, thus incurring a higher cost due to the unit cost premium of MAs. Even at moderate sizes, the limited spatial degrees of freedom prevent the MA position optimization from sufficiently reducing the RF chain count to offset this premium. Consequently, the FPA architecture is more cost-effective for compact regions, while the proposed MA scheme is superior for realizing large-aperture transmitters by exploiting spatial diversity to minimize hardware usage.

\subsubsection{Impact of Number of Target Areas}
Fig.~\ref{Fig:cost_vs_J} depicts the average normalized deployment cost versus the number of target areas $J$. A significant performance gap emerges between the proposed joint MA-IRS scheme and the per-area union benchmark as $J$ increases. At $J=1$, the two schemes yield identical costs since the resource union for a single area is trivial. However, as $J$ grows, the cost of the per-area union scheme rises rapidly. This is because it optimizes resources for each area independently and aggregates them without exploiting potential overlaps. In contrast, the cost of the proposed joint MA-IRS scheme increases only marginally. This highlights the advantage of resource reuse. By identifying a minimum common set of hardware reconfigurable to serve different areas, the proposed joint MA-IRS scheme avoids the redundancy of dedicated deployments. Meanwhile, the FPA-IRS and MA-(all IRS) schemes incur much higher costs due to the deployment of excessive fixed resources which are insensitive to the actual coverage requirements. Finally, as shown in the zoomed-in view, the proposed element pruning scheme consistently achieves the lowest cost by eliminating redundant elements from the selected IRSs. 

\begin{figure}[!t]
	\vspace{-3mm}
	\centering
	\includegraphics[scale=0.65]{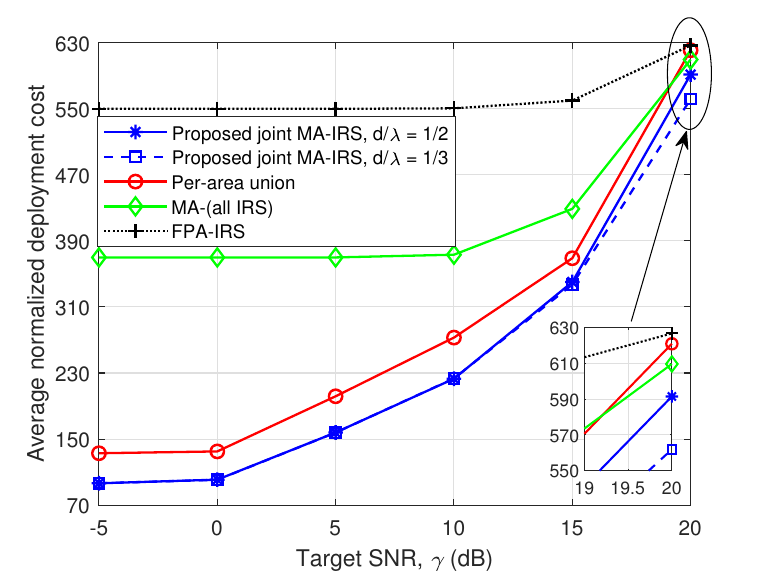}
	\caption{Average normalized deployment cost versus target SNR.} \label{Fig:cost_vs_SNR}
\end{figure}

\begin{figure}[!t]
	\vspace{-3mm}
	\centering
	\includegraphics[scale=0.67]{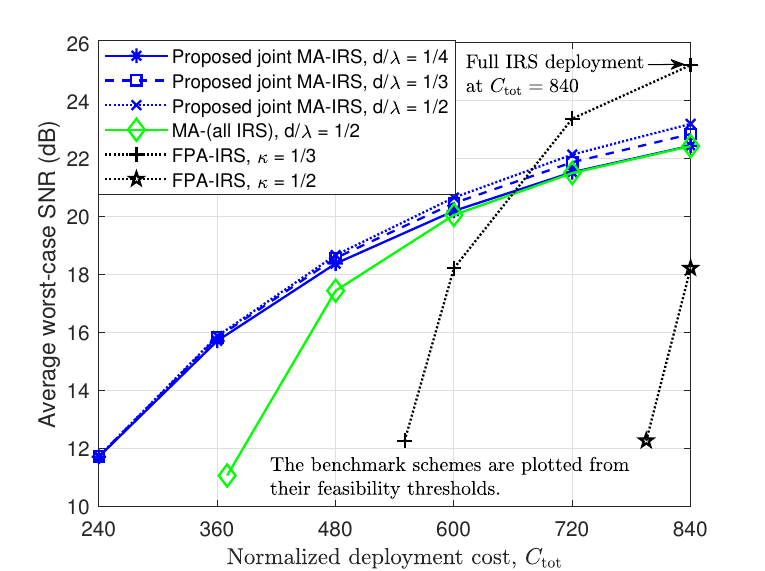}
	\caption{Average worst-case SNR versus normalized deployment cost.} \label{Fig:SNR_vs_cost}
	\vspace{-3mm}
\end{figure}

\subsubsection{Impact of Target SNR}
Fig.~\ref{Fig:cost_vs_SNR} presents the average normalized deployment cost versus the target SNR $\gamma$. As anticipated, the deployment costs for all schemes escalate as $\gamma$ increases since more active MAs and passive IRSs are required to meet the stricter QoS requirements. Throughout the plotted range, the proposed joint MA-IRS scheme consistently achieves the lowest cost among all considered benchmarks.

Two interesting phenomena are observed in the high-SNR regime at $\gamma=20$~dB. First, the cost gap between the proposed joint MA-IRS scheme and the MA-(all IRS) benchmark becomes much smaller. This is because the stringent SNR target forces the proposed algorithm to activate most candidate IRSs. As the active-IRS set approaches the full set, the cost-saving potential from IRS selection largely vanishes, and the proposed scheme effectively converges to the all-IRS benchmark. Second, the proposed scheme with $d/\lambda=1/3$ achieves a lower cost than that with $d/\lambda=1/2$. Although this seems counter-intuitive given the smaller allowable MA count for $d/\lambda=1/3$ in Fig.~\ref{Fig:feasibility_vs_d} (i.e., $M_{\max} = 25$ versus $49$), the key is that at $\gamma=20$~dB the required MA number remains below $M_{\max}$ for $d/\lambda=1/3$. In this non-saturated regime, the $d/\lambda=1/3$ grid provides finer spatial resolution than $d/\lambda=1/2$, enabling more effective exploitation of small-scale fading and higher end-to-end channel gains. Consequently, the target SNR can be met with lower hardware expenditure. By contrast, in the low-SNR regime, the hardware requirement is already minimal, and the additional spatial gain from a finer grid is insufficient to reduce the discrete integer numbers of deployed elements, leading to negligible cost differences between the two step sizes.

\vspace{-3mm}
\subsection{Worst-Case SNR Maximization Under a Cost Budget}
Fig.~\ref{Fig:SNR_vs_cost} depicts the average worst-case SNR versus the normalized deployment cost budget, highlighting the trade-off between MA flexibility and the low unit cost of FPAs. 

In the low-to-medium cost regime, the proposed joint MA-IRS scheme consistently achieves a higher worst-case SNR than all benchmarks under the same budget. In particular, the FPA-IRS benchmark is infeasible when $C_{\rm tot}<550$ for $\kappa=1/3$ and when $C_{\rm tot}<795$ for $\kappa=1/2$, indicating a high feasibility threshold under the adopted FPA deployment setting. By contrast, MA-based designs exploit position optimization to harvest spatial diversity with only a few active RF chains, making them attractive under tight budget constraints. 

In the high-cost regime, relative performance depends heavily on the unit cost ratio $\kappa$. When FPAs are significantly cheaper with $\kappa=1/3$, the FPA-IRS benchmark reaches approximately $25$~dB at $C_{\rm tot}=840$ by enabling full deployment. This allows the system to approach the aperture-limited performance ceiling. In contrast, the proposed MA-IRS scheme optimizes both MAs and IRSs under the same budget constraint. Due to the higher unit cost of MAs, the restricted budget limits the number of deployable MAs and IRSs compared to the fully populated FPA benchmark. Consequently, this trade-off may yield a slightly lower SNR despite the benefits of position optimization. However, with moderately expensive FPAs ($\kappa=1/2$), the MA scheme retains its superiority even at high budgets. Furthermore, among the proposed schemes, the grid with $d/\lambda=1/3$ outperforms $d/\lambda=1/2$. This confirms that in the non-saturated regime where the number of MAs is far below $M_{\max}$, a finer grid resolution provides higher spatial degrees of freedom to improve beamforming gain. \looseness=-1

\section{Conclusion}\label{sec:_conclusion}
In this paper, we proposed a two-scale spatial deployment framework where macroscopic IRS site selection shapes the propagation geometry and microscopic MA repositioning exploits local channel variations. We formulated a joint optimization problem to minimize the total hardware cost by coordinating IRS site selection, MA positioning, and beamforming design, subject to specific QoS requirements for multiple target areas. A penalty-based AO algorithm was first devised to solve the feasibility and cost minimization problems, followed by a secondary element refinement stage to strictly eliminate hardware redundancy. Simulation results validated the effectiveness of the proposed designs and provided critical engineering insights into MA grid discretization, as well as guidelines for selecting between MA and fully populated FPA architectures depending on placement aperture sizes for cost minimization, and budget levels for worst-case SNR maximization.

\bibliographystyle{IEEEtran}
\bibliography{ref}

\end{document}